\documentclass[reprint,nofootinbib]{revtex4-1}
\usepackage{graphicx}
\usepackage{aas_macros}

\usepackage{amsmath}
\graphicspath{{./figures/}}

\newcommand{\eq}[1]{\begin{equation}\begin{split}#1\end{split}\end{equation}}
\newcommand{\ea}[1]{\begin{align}#1\end{align}}
\def\bxi{{\mbox{\boldmath $\xi$}}}
\def\br{{\bf r}}
\def\bv{{\bf v}}
\def\bOmega{{\bf \Omega}}
\def\Oms{\Omega_s}
\def\Omo{\Omega_{\rm orb}}
\def\sigi{\sigma_\alpha}
\def\omi{\omega_\alpha}
\def\jm{j_{\rm max}}
\def\al{_\alpha}
\def\be{\begin{equation}}
\def\ee{\end{equation}}

\begin{document}
\title{Resonant Tidal Excitation of Oscillation Modes in Merging Binary Neutron Stars: 
Inertial-Gravity Modes}
\author{Wenrui Xu and Dong Lai}
\affiliation{Cornell Center for Astrophysics and Planetary Science, Cornell University, 
Ithaca, NY 14853, USA}

\begin{abstract}
In coalescing neutron star (NS) binaries, tidal force can resonantly
excite low-frequency ($\lesssim 500$~Hz) oscillation modes in the NS,
transferring energy between the orbit and the NS. This resonant tide can
induce phase shift in the gravitational waveforms, and potentially
provide a new window of studying NS interior using gravitational
waves. Previous works have considered tidal excitations of pure g-modes
(due to stable stratification of the star) and pure inertial modes
(due to Coriolis force), with the rotational effect treated in an
approximate manner. However, for realistic NSs, the buoyancy and
rotational effects can be comparable, giving rise to mixed
inertial-gravity modes. We develop a non-perturbative numerical
spectral code to compute the frequencies and tidal coupling
coefficients of these modes. We then calculate the phase shift in the
gravitational waveform due to each resonance during binary inspiral.
Given the uncertainties in the NS equation of state and stratification
property, we adopt polytropic NS models with a parameterized
stratification. We derive relevant scaling relations and survey
how the phase shift depends on various properties of the NS. We find that
for canonical NSs (with mass $M=1.4M_\odot$ and radius $R=10$~km) and
modest rotation rates ($\lesssim 300$~Hz), the gravitational wave
phase shift due to a resonance is generally less than 0.01 radian.
But the phase shift is a strong function of $R$ and $M$, and can reach
a radian or more for low-mass NSs with larger radii ($R\gtrsim
15$~km). Significant phase shift can also be produced when the
combination of stratification and rotation gives rise to a very low
frequency ($\lesssim 20$~Hz in the inertial frame) 
modified g-mode. As a by-product of our precise
calculation of oscillation modes in rotating NSs, we find that some
inertial modes can be strongly affected by stratification; we also
find that the $m=1$ r-mode, previously identified to have a small but
finite inertial-frame frequency based on the Cowling approximation,
in fact has essentially zero frequency, and therefore cannot be excited during the 
inspiral phase of NS binaries.
\end{abstract}

\maketitle

\section{Introduction}

The recent breakthrough in the detection of gravitational waves (GWs)
from merging black hole (BH) binaries by advanced LIGO \cite{Abbottetal16a,Abbottetal16b,Abbottetal17} heralds a new era of studying
compact objects using GWs. Coalescing neutron star-neutron star
(NS-NS) and NS-BH binaries have long been considered the most
promising sources of GWs for LIGO/VIRGO
\cite{Cutleretal93,CutlerThorne02}.  The last few minutes of the
binary inspiral produce GWs with frequencies sweeping upward through
the LIGO sensitivity band (10-1000~Hz).  Due to the expected low
signal-to-noise ratios, accurate gravitational waveforms are required
to serve as theoretical templates that can be used in matched
filtering to detect the GW signal from the noise and to extract binary
parameters from the waveform. 

The possibility of using GWs from NS binaries to constrain the
equation of state (EOS) of dense nuclear matter has long been
recognized \cite{Cutleretal93}.  The gravitational waveforms
associated the final merger of two NSs or the tidal disruption of a NS
by a BH exhibit power spectra with characteristic frequencies that
reflect the dynamical frequency of the NS, $(GM/R^3)^{1/2}$ (where
$M,~R$ are the NS mass and radius); these characteristic frequencies
can be used to constrain the NS radius and thus the EOS (given the
mass measurement from the inspiral waveform) (e.g., \cite{BildstenCutler92,LaiWiseman96,Shibataetal05,Bausweinetal14,Foucartetal14,Foucartetal16}).
Since these characteristic frequencies are greater than kHz, beyond
the current aLIGO sensitivity band, measuring them will be challenging
without special experimental effort to enhance the high-frequency
sensitivity of the LIGO interferometer.

\subsection{Quasi-Equilibirum Tides}

Another method to constrain the EOS of NSs is to use tidal effect.
Numerous papers have been written on the effect of quasi-equilibirum
tides on the inspiral waveforms.  The quasi-equilibrium tide
corresponds to the global (f-mode), quadrupolar deformation of the NS. To the
leading (Newtonian) order, this tidal deformation changes the
interaction potential between the two stars (with the NS mass $M$ and
radius $R$, the companion mass $M'$ -- treated as a point mass)
from $V^{(0)}(r)=-GMM'/a$ 
(where $a$ is the binary separation) to
\begin{equation}
V(r)=-{GMM'\over a}-{\cal O}\left({k_2G{M'}^2R^5\over a^6}\right),
\end{equation}
where $k_2$ is the so-called Love number.
This leads to a correction to the GW phase (``phase shift'')
\begin{equation}
d\Phi= d\Phi^{(0)}\left[1-{\cal O}\left({k_2M'R^5\over Ma^5}\right)\right],
\label{eq:dPhi}\end{equation}
with the ``point-mass'' GW phase given by 
\begin{equation}
d\Phi^{(0)}={5\over 48(\pi M_c f)^{5/3}} d\ln f,
\end{equation}
where $f$ is the GW frequency and $M_c=(MM')^{3/5}/(M+M')^{1/5}$ is the chirp mass.
For Newtonian polytropic NS models, simple analytic expressions
(including the effect of rotation) for the phase shift are given in
Ref.~\cite{Laietal94} (Eqs.~66 and 72; see also \cite{Kochanek92}).
Semi-analytic GR calculations of such
quasi-equilibrium tidal effect (including more precise determination
of the Love number) can be found in numerous papers (e.g.,
\cite{FlanaganHinderer08, BinningtonPoisson09, DamourNagar09, Penneretal11, Ferrarietal12}). Obviously
this effect is only important at small orbital separations, just prior
to merger. Again, there is some prospect of measuring this, thereby
constraining the EOS, but it will be challenging because of the limited 
high-frequency sensitivity of aLIGO \cite{Damouretal12}.  At small orbital separations, there is also a ``dynamical''
correction to the above ``equilibrium'' phase shift expression
\cite{Hindereretal16}. This arises from the finite response time $\omega_{\rm f}^{-1}$
of the NS (where $\omega_{\rm f}$ the quadrupole f-mode frequency)
as compared to the tidal forcing time $\omega_{\rm tide}^{-1}$
(where $\omega_{\rm tide}=2\Omega$ for nonrotating NSs, with $\Omega$ the orbital frequency)
\cite{Lai94}. This ``dynamical'' correction essentially amounts to replacing $k_2$ by
$k_2/(1-4\Omega^2/\omega_{\rm f}^2)$. Thus
Equation (\ref{eq:dPhi}) becomes
\begin{equation}
d\Phi= d\Phi^{(0)}\left[1-{\cal O}\left({k_2M'R^5\over Ma^5}\right)
{1\over 1-4\Omega^2/\omega_{\rm f}^2}\right].
\label{eq:dPhi2}
\end{equation}
Finally, we note that the quadrupole approximation is not accurate at small orbital separations,
and one must use numerically computed quasi-equilibrium binary
sequences to characterize the full tidal effects \cite{Baumgarteetal98,Uryuetal09} 
or use 3D hydrodynamical simulations (e.g. \cite{Shibataetal05,Foucartetal16}).

\subsection{Resonant (Dynamical) Tides}  

In the early stage of the inspiral, with the GW frequencies between
$10$~Hz to a few hundred Hz, it is commonly assumed that a NS can be
treated as a point mass, and tidal effects are completely
negligible. This is indeed the case for the quasi-equilibrium tides
discussed above. However, a NS can possess a variety of low-frequency
($\lesssim 500$~Hz) oscillation modes due to stable density
stratification and/or rotation.  During binary inspiral, the orbit can
momentarily come into resonance with the normal modes of the NS.  By
drawing energy from the orbit and resonantly exciting the modes, the
inspiral speeds up around the resonant frequency, giving rise to a
phase shift in the GW.  This problem was first studied in the case of
non-rotating NSs \cite{ReiseneggerGoldreich94,Lai94,Shibata94}
where the only modes that can be resonantly excited are g-modes, with
typical mode frequencies $\alt 100$~Hz
(\cite{ReiseneggerGoldreich94,Lai94} considered g-modes associated
with the bulk composition gradients, while \cite{Shibata94}
considered those associated with crustal density jumps).  It was found
that the effect is small for typical NS parameters (mass $M\simeq
1.4M_\odot$ and radius $R\simeq 10$~km) and several equations of state
\cite{Lai94} because the coupling between the g-mode and the tidal
potential is weak. Superfluidity in the NS core can significantly
affect the g-mode property \cite{KantorGusakov14,Passamontietal16}, and
recent studies suggest that the resulting phase shift has the same
order of magnitude as that of a normal fluid NS
\cite{YuWeinberg17a,YuWeinberg17b}.

Ho \& Lai (1999)\cite{HoLai99} studied the effect of NS rotation, and
found that the g-mode resonance can be strongly enhanced even by a
modest rotation (e.g., the phase shift in the waveform $\Delta\Phi$
reaches up to $0.1$~radian for a spin frequency $\nu_s\alt 100$~Hz) because rotation can reduce the g-mode frequency.
For a rapidly rotating NS ($\nu_s\agt 500$~Hz), f-mode resonance
becomes possible (since the inertial-frame f-mode frequency can be
significantly reduced by rotation) and produces a large ($\gg 1$)
phase shift. They also studied the Coriolis-force driven r-modes, and
found that their tidal excitations become appreciable for very rapid
NS rotations. These r-modes can also be excited by (post-Newtonian)
gravitomagnetic force \cite{FlanaganRacine07}, with the resulting GW
phase shift comparable to the Newtonian resonant tidal excitation.

A rotating NS supports a large number of Coriolis-force driven modes
named inertial modes (i-modes, also called rotational hybrid modes or
generalized r-modes; see, e.g.,
Refs.~\cite{PapaloizouPringle81,LockitchFriedman99,
  Schenketal02,Wu05,Passamontietal09}), of which r-mode is a member.
Most i-modes have frequencies of order the NS spin frequency. Based on
approximate calculations, Lai \& Wu (2006) \cite{LaiWu06} found that
i-modes have coupling to the Newtonian tidal potential similar to the
r-modes, and most i-mode tidal resonances give relatively small GW
phase shifts. They also identified one r-mode that has a 
rather small 
inertial-frame frequency, which implies a large phase shift
independent of the spin frequency. This mode gives a GW frequency that
is typically (for reasonable NS rotations) below the aLIGO sensitivity
band (see Section IV.B.4 for our new result on this mode).

Other related studies include tidal excitation of shear modes associated with NS
crusts \cite{Tsangetal12} (producing small/modest phase shift) 
and possible nonlinear effects due to the coupling of ``off-shell'' f-modes to
high-order g-modes and p-modes \cite{Weinbergetal13,Venumadhavetal14,
Weinberg16} -- we will not study these issues in this paper.

\subsection{This Paper}

Overall, previous studies (reviewed above) suggest that for astrophysically most likely
NS parameters ($M\simeq 1.4M_\odot$, $R\simeq 10$~km, $\nu_s\alt
100$~Hz), tidal resonances have a small effect on the gravitational
waveform during binary inspiral (with the GW phase shift $\Delta\Phi\ll 1$).
However, it is important to keep in mind that the effect is a strong function of $R$,
and a larger NS radius ($R\simeq 15$~km), appropriate for $\sim 1.4M_\odot$ NSs with 
stiff EOS or low-mass ($\lesssim 1M_\odot$) NSs,
would significantly increase the phase shift. In the case of g-modes, the magnitude
of $\Delta\Phi$ depends on several uncertain aspects of nuclear EOS
(e.g. the symmetry energy). Although the observed double NS systems all have rather modest 
rotation rates ($\lesssim 50$~Hz), rapidly rotating NSs ($\sim 700$~Hz) have 
been found. Future GW observations may reveal new classes of NSs
that are totally different from those already observed via electromagnetic radiation.
To this end, it is desirable to examine tidal resonances for a wide range of 
NS parameters and survey various possibilities. This is one of the main goals 
of this paper.

Previous studies of inertial mode resonances during binary inspiral
have adopted approximate calculation of these modes. Given that the GW
phase shift due to resonance depends on the tidal coupling
coefficient, which in turn depends sensitively on the shape of the
mode wavefunction, there is a concern that an inaccurate treatment may
lead to large error.  Indeed, we show in this paper that one of the
``important'' r-modes identified in Ref.~\cite{LaiWu06} turns out to
have identically zero frequency (see Section IV.B). Moreover,
previous calculations of the mixed ``inertial-gravity'' modes of NSs
are not accurate, particularly in the regime where stratification and
Coriolis force are comparable -- and it is precisely in this regime
(where the mode frequency is close to zero in the inertial frame) that
a significant phase shift is expected. Note that accurate calculations
of mixed modes in main-sequence stars \cite{Ballotetal10} and NSs
\cite{Passamontietal09} do exist (the latter use an initial-value
problem to determine mode frequencies but not eigenfunctions), but
they do not calculate the tidal coupling coefficients of the modes.
In this paper, we develop a new spectral code to calculate the
inertial-gravity modes (both the frequency and tidal coupling
coefficient) precisely, including the full treatment of Coriolis force, gravitational potential
perturbation and the effect of rotational distortion,
and we use the results to evaluate the
significance of tidal resonances of stratified, rotating NSs.

Our paper is organized as follows. Section II summarizes the key equations
for calculating the gravitational wave phase shift due to tidal resonance.
Section III describes our method for numerical computation of the oscillation modes
of rotating, stratified NSs. In Section IV we discuss the key results of NS oscillation
modes (including scaling relations) that directly influence resonant tidal
excitation. We present our results for the GW phase shifts associated with
various mode resonances in Section V and conclude in Section VI.

\section{Tidal resonance during binary inspiral}

The method of calculating the GW phase shift due to tidal resonance in
a rotating NS with arbitrary spin-orbit misalignment was presented in
Refs.~\cite{HoLai99,LaiWu06}.  Here we introduce the notations and
give the key equations.

Consider a NS of mass $M$, radius $R$ and spin $\bOmega_s$ in orbit
with a companion of mass $M'$ (another NS or a black hole).  We allow
for a general spin-orbit inclination angle $\Theta$ (the angle between
$\bOmega_s$ and the orbital angular momentum ${\bf L}$).  The orbital
radius $a$ decreases and the orbital angular frequency $\Omo$
increases in time due to GW emission.
In the spherical coordinate system centered on $M$ with the $Z$-axis along ${\bf L}$,
the gravitational potential produced by $M'$ is (to quadrupole order):
\eq{
U(\br,t) = &-\frac{GM'r^2}{a^3}\left(\frac{3\pi}{10}\right)^{1/2}\left[
e^{-2i\Phi_{\rm orb}(t)}Y_{22}(\theta_L,\phi_L)\right. \\
&\left.+ e^{2i\Phi_{\rm orb}(t)}Y_{2,-2}(\theta_L,\phi_L)\right],
}
where $\Phi_{\rm orb}(t)=\int^t dt\,\Omo$ is the orbital phase. We ignore
higher order components of the tidal potential since they have little
contribution to the tidal coupling.

In order to describe oscillation modes relative to 
the spin axis, we express the tidal potential in terms of 
$Y_{lm}(\theta,\phi)$, the spherical harmonic function defined in 
the corotating frame of the NS with the $z$-axis along $\bOmega_s$. 
This is achieved by the relation
\eq{
Y_{2m'}(\theta_L,\phi_L)=\sum_{m}{\cal D}^{(2)}_{mm'}
(\Theta)Y_{2m}(\theta,\phi_s),
}
where ${\cal D}^{(2)}_{mm'}$ is the Wigner ${\cal D}$-function and $\phi_s=\phi+\Oms t$.

Oscillation modes of the NS are specified by the Lagrangian
displacement, $\bxi(\br,t)$, of a fluid element from its unperturbed
position.  In the rotating frame, a free mode of frequency
$\omega_\alpha$ has
$\bxi_\alpha(\br,t)=\bxi_\alpha(\br)\,e^{-i\omega_\alpha t}\propto
e^{im\phi-i\omega_\alpha t}$, where $m$ is the azimuthal number of the
mode and $\alpha$ denotes the mode index (which includes $m$). We only
need to consider $m>0$, since a mode with $(m,\omega_\alpha)$ is physically
identical to a mode with $(-m,-\omega_\alpha)$. In this convention ($m>0$), 
a mode with $\omega_\alpha>0$ ($\omega_\alpha<0$) is prograde (retrograde) with respect to 
the rotation.

A tidal resonance occurs when a mode with the inertial-frame frequency
\be
\sigi = \omi + m\Oms,
\ee
is excited by the potential component $\propto e^{-im'\Phi_{\rm orb}}$,
with the mode frequency satisfying the condition
\eq{
\sigi = m'\Omo.
}
Note that for quadrupolar tide, we only need to consider $m' = \pm 2$. 
Clearly, a prograde mode ($\sigma_\alpha>0$) is excited by the $m'=2$ potential, 
while a retrograde mode ($\sigma_\alpha<0$) by the $m'=-2$ potential.
The energy transferred to the mode during the resonance is given by
\eq{
\Delta E_{\alpha,m'}=&\frac{3\pi}{10}{G{M'}^2\over R}{GM\over R^3}
\left({\pi\over m'\dot\Omega_{\rm orb}}\right)
{\sigma_\alpha\over\varepsilon_\alpha}\\
& \times \left({\cal D}_{mm'}^{(2)}Q_{\alpha,2m}\right)^2
\left({R\over a_\alpha}\right)^{6},
\label{eq:deltaE}
}
where $\dot\Omega_{\rm orb}$ is the rate of change of $\Omo$ due to GW
emission, $a_\alpha$ is the binary semi-major axis at the tidal resonance, and
\ea{
&Q_{\alpha,2m}\equiv \bigl\langle\bxi_\alpha,\nabla (r^2Y_{2m})\bigr\rangle,
\label{eq:Qdefine}\\
& \varepsilon_\alpha\equiv
\omi+\langle\bxi_\alpha,i\bOmega_s\times\bxi_{\alpha}\rangle,
}
with $\langle A,B\rangle \equiv \int d^3x\rho(A^*\cdot B)$, and we use the normalization 
$M=R=1$ and $\langle\bxi_\alpha,\bxi_\alpha\rangle=1$. 
The quantity $Q_{\alpha,2m}$ is a nondimensional number characterizing
the strength of the tidal coupling of the mode.  The phase shift in GW
signal caused by this energy transfer is given by
\eq{
\Delta\Phi=&-\frac{5\pi^2}{1024}\left(\frac{Rc^2}{GM}
\right)^5\frac{1}{q(1+q)}\\
& \times\frac{m'}{\hat \varepsilon_\alpha|\hat\sigma_\alpha|}
\left({\cal D}^{(2)}_{mm'}
Q_{\alpha,2m}\right)^2, \label{eq:orbchange}
}
where $\hat\sigma_\alpha=\sigma_\alpha (R^3/GM)^{1/2}$
and $\hat\varepsilon_\alpha=\varepsilon_\alpha (R^3/GM)^{1/2}$.
Note that in the above equation, $m'=\pm 2$ and $m=1$ or $2$;
modes with larger $m$ do not couple with the quadrupole tidal potential.

\section{Modes in Rotating Neutron Stars: Method of Calculation}

This section describes our numerical method to calculate the
oscillation modes of rotating, stratified neutron stars.  For
simplicity, we assume the equilibrium state of the NS to be barotropic
and spherically symmetric, i.e., the centrifugal distortion due to
rotation is ignored; this should not affect our lowest order results
when $\Oms$ is relatively small compared to $(GM/R^3)^{1/2}$. (For one
particular inertial mode, the effect of centrifugal distortion can be
important -- this will be discussed in Section IV.B). We include the
full effects of the Coriolis force and gravitational potential
perturbation in the mode calculation.

The numeric code we use here is based on the spectral method developed
by Reese et al.~(2006) \cite{Reeseetal06} to calculate p-modes in
rapidly rotating and stratified stars with polytropic density profile and
stratification characterized by a constant adiabatic exponent. 
We modify this method to allow for general density and stratification
profiles. Reese et al. also included the distortion of the star due to
rotation in their calculation. For most part of this paper we ignore
this distortion for simplicity, but its effect can be incorporated in
our method without much technical difficulty (see Section IV.B.2).

Let the equilibrium (unperturbed) stellar profile be given by density
$\rho_0(r)$, pressure $p_0(r)$ and gravitational potential
$\Psi_0(r)$. In the rotating frame, the Eulerian perturbation of
density, pressure and gravitational potential are denoted by
$\delta\rho,\delta p$ and $\delta\Psi$, respectively. The velocity
perturbation $\delta\bv$ is related to the Lagrangian displacement
$\bxi$ by $\delta\bv = -i\omega\bxi$, where we have assumed 
$\bxi\propto e^{-i\omega t}$ and $\omega$ is the mode frequency in
the rotating frame. The linearized fluid dynamics equations then reduce to 
a generalized eigenvalue problem:
\ea{
-i\omega\delta\rho &= -\nabla\cdot(\rho_0\delta\bv),\label{unscaled1}\\
-i\omega\rho_0\delta\bv &= -\nabla \delta p +\delta\rho \mathbf{g}_0 - \rho_0\nabla\delta\Psi - 2\rho_0\mathbf{\Oms}\times\delta\bv,\\
-i\omega(\delta p - c_0^2\delta \rho) &= \frac{\rho_0N_0^2c_0^2}{||\mathbf{g}_0||^2}\delta\bv\cdot \mathbf{g}_0,\\
0 &= \Delta\delta\Psi - 4\pi G\delta\rho.\label{unscaled4}
}
The gravitational acceleration $\mathbf{g}_0$, addiabatic sound speed $c_0$
and Brunt-V\"ais\"al\"a frequency $N_0$ are given by
\ea{
&\mathbf{g}_0 = -\nabla\Psi_0,\\
&c_0^2 = \Gamma\frac{p_0}{\rho_0},\\
&N_0^2 = \mathbf{g}_0\cdot\frac{\nabla\rho_0}{\rho_0}\left(1-\frac{\gamma}{\Gamma}\right),
}
where $\gamma$ and $\Gamma$ are defined by
\ea{
\gamma&\equiv \frac{d\ln p_0}{d\ln\rho_0},\\
\Gamma&\equiv \left(\frac{\partial\ln p}{\partial\ln\rho}\right)_{\rm ad}.
}
The subscript ``ad'' denotes adiabatic derivative. In general $\Gamma$ and $\gamma$
depend on $r$. When the star has a polytropic density profile,
$\gamma$ is constant and corresponds to the polytropic exponent.

To solve the eigenvalue problem,
we decompose $\delta\bv$ into spheroidal and toroidal components:
\eq{
\delta\bv = &\sum_{j=m}^{\infty}\biggl[ u_{jm}Y_{jm}\mathbf{e}_r +
  v_{jm}\left(\partial_\theta Y_{jm}\mathbf{e}_\theta + D_\phi
  Y_{jm}\mathbf{e}_\phi\right) \\&+ w_{jm}
\left(D_\phi Y_{jm}\mathbf{e}_\theta - \partial_\theta
  Y_{jm}\mathbf{e}_\phi\right)\biggr],
\label{uvw_def}}
where $u_{jm},v_{jm},w_{jm}$ are functions of $r$,
$\mathbf{e}_r,\mathbf{e}_\theta,\mathbf{e}_\phi$ are unit vectors in the
$r,\theta,\phi$ directions, respectively, and $D_\phi\equiv
(\sin\theta)^{-1}\partial_\phi$. The terms with $u_{jm},v_{jm}$ give the spheroidal
component of the velocity perturbation and the terms with $w_{jm}$ the toroidal
component.

More details of our spectral algorithm are given in the Appendix (see
also Ref.~\cite{Reeseetal06}), where we also discuss the limitation of the method.

\section{Oscillation Modes and tidal coupling coefficients}

There have been numerous theoretical studies of NS oscillations,
taking account of ``realistic" NS structure and equation of state
(including stratification, superfludity, crustal rigidity, etc.) and
general relativity (e.g., \cite{McDermottetal88, Krugeretal15}). While
for non-rotating NSs, the computation of various non-radial modes (for
a given NS model) can be achieved efficiently, for rotating NSs, there
remain appreciable technical challenges to compute oscillation modes
in the non-perturbative regime (e.g.,
\cite{Passamontietal09,Donevaetal13}; see
\cite{PaschalidisStergioulas16} for a review). Given the uncertainty
in the NS interior equation of state,  in this paper we mainly consider Newtonian
polytropic models (both in terms of the density profile and
stratification), in order to provide a survey of different
possibilities and to identify the most important oscillation modes for
tidal resonance.

\subsection{Pure g-Modes}

\subsubsection{G-mode Scaling Relations}

For non-rotating NSs ($\Omega_s=0$), only gravity modes (g-modes) have
sufficiently low frequencies to allow resonant excitation during
binary inspiral. G-modes in NSs arise from composition (e.g. proton to
neutron ratio) gradient in the stellar core
\cite{ReiseneggerGoldreich92}, density discontinuities in the crust
\cite{Finn87,Strohmayer93} as well as thermal buoyancy associated with
finite temperatures \cite{McDermottetal88}. For cold NSs (as expected
in merging binaries), the composition gradient in the core provides
the strongest restoring force. The core g-modes are
sensitive to the symmetry energy of nuclear matter \cite{Lai94}, and
are also affected by the presence of superfluidity (in which case the
restoring force for g-modes arise from the gradient in
muon-to-electron fraction; see
\cite{KantorGusakov14,Passamontietal16}).  Here for simplicity we
first consider polytropic NS models with constant $\gamma = 2$ and
constant adiabatic index $\Gamma>\gamma$. We vary $\Gamma$ to survey
different strength of stratification. This model should give a
relatively good estimation of the frequency and tidal coupling of
g-modes in realistic NSs.

G-modes are purely spheroidal [i.e. $w_{jm}=0$ in Eq.~\eqref{uvw_def}] and
has only one nonzero $j$ component. Thus we can label a mode by three
numbers, $j,m$ and $n$, where $n$ corresponds to the number of nodes in
$u_{jm}$. Because of the symmetry, the mode frequency and tidal
coupling coefficient depend only on $j$ and $n$. 
Only $j=2$ modes couple to the quadrupole ($l=2$) tidal potential.

\begin{table*}
\centering
\caption{Scaled frequency and tidal coupling coefficient\footnote{The scaled quantities are defined 
by Eqs.~(\ref{scale_g_1})-(\ref{scale_g_2}).} for $j=2$
pure g-modes in NSs with different density profiles, assuming constant
  $\Gamma-\gamma$.}
\label{g_modes}
\begin{ruledtabular}
\begin{tabular}{c@{\hskip .5in}cc@{\hskip .5in}cc@{\hskip .5in}cc}
~&\multicolumn{2}{c@{\hskip .5in}}{$\gamma=1.5$} & \multicolumn{2}{c@{\hskip .5in}}{$\gamma=2$} & \multicolumn{2}{c}{$\gamma\simeq3$\footnote{This model has $\gamma\simeq 3$ except for a thin layer near the surface 
where $\gamma$ drops to 2. 
For details of this model see Section IV.A.2.}} \\
~	&$\bar{\omega}\al$ $(2\pi{\rm Hz})$     & $\bar{Q}_{\alpha,2m}$      & $\bar{\omega}\al$ $(2\pi{\rm Hz})$   & $\bar{Q}_{\alpha,2m}$     & $\bar{\omega}\al$ $(2\pi{\rm Hz})$   & $\bar{Q}_{\alpha,2m}$      \\\hline
$n=1$	& $\pm$429 & $1.2\times 10^{-3}$	&$\pm$181.3 & $3.5\times 10^{-4}$	& $\pm$92.8 & $7.5\times 10^{-5}$ \\
$n=2$	& $\pm$309 & $5.2\times 10^{-4}$	&$\pm$124.6 & $8.5\times 10^{-5}$	& $\pm$61.7 & $1.2\times 10^{-5}$ \\
$n=3$	& $\pm$242 & $2.4\times 10^{-4}$	&$\pm$95.7  & $2.5\times 10^{-5}$	& $\pm$46.9 & $3\times 10^{-6}$
\end{tabular}
\end{ruledtabular}
\end{table*}

The middle two columns in Table \ref{g_modes} give the scaled 
frequency and tidal coupling coefficient of $j=2$ g-modes with
different $n$ for a $\gamma=2$ polytropic NS. Our numerical results
show that (to lowest order)
\ea{
&\omega\al \propto (\Gamma-\gamma)^{1/2}M^{1/2}R^{-3/2},\\
&Q_{\alpha,2m} \propto \Gamma-\gamma,
}
so we define scaled frequency and tidal coupling coefficient by
\ea{
&\bar\omega\al = \omega\al\left(\frac{\Gamma-\gamma}{0.01}\right)^{-1/2}M_{1.4}^{-1/2}R_{10}^{3/2},\label{scale_g_1}\\
&\bar Q_{\alpha,2m} = Q_{\alpha,2m}\left(\frac{\Gamma-\gamma}{0.01}\right)^{-1},\label{scale_g_2}
}
where $M_{1.4} = M/(1.4M_\odot)$ and $R_{10}=R/(10\,{\rm km})$. This scaling
ensures that the results are independent of the NS mass, radius and
stratification ($\Gamma-\gamma$) when $\Gamma-\gamma \ll 1$.  Note
that the scaled variables are equal to the physical frequency and tidal
coupling coefficient for a canonical NS with $M=1.4M_\odot, R=10$~km
and $\Gamma-\gamma=0.01$. The $(\Gamma-\gamma)$ scaling of $\omega\al$
we obtain agrees with the analytical WKB estimate of the mode frequency
(e.g. \cite{Lai94})
\eq{
\omega\al \simeq \frac{\sqrt{j(j+1)}}{(n+C)\pi}\int_0^R\frac{N_0(r)}{r}dr,
}
where $C$ is a constant of order unity, and 
the Brunt-V\"ais\"al\"a frequency $N_0(r)$ is given by 
\be
N_0(r)=g_0 \sqrt{\rho_0\over p_0}\left({1\over\gamma}-{1\over\Gamma}\right)^{1/2},
\ee
which is proportional to $\left(\Gamma-\gamma\right)^{1/2}$ for $(\Gamma-\gamma)\ll 1$.

We see from Table \ref{g_modes} that $Q_{\alpha,2m}$ decreases
significantly with increasing $n$; therefore the $n=1$ g-modes are the
only modes that contribute significantly to the phase shift induced by
tidal resonances. Note that for realistic NS stratification
($\Gamma-\gamma\sim0.01$), the tidal coupling is small even fore
the $n=1$ modes.

\subsubsection{Effect of Different Stellar Density Profiles}

Canonical NSs with mass around $1.4M_\odot$ 
can be approximated by a polytrope with $\gamma$ between 2 and 3,
and low-mass ($\lesssim 1M_\odot$) could be modeled with $\gamma=1.5$.
Therefore, it is useful to investigate how different NS
density profiles (different $\gamma$) affect the g-modes. Table
\ref{g_modes} contains the scaled frequency and tidal coupling coefficient
of g-modes for two other NS models. One of them is a $\gamma=1.5$
polytrope. The other has 
\be
\gamma(r) = 2+\tanh\left[{5\rho_0(r)\over\rho_c}\right], 
\ee
with $\rho_c$ the central density of the NS;
this model has $\gamma\simeq 3$ except for a thin layer near the
surface where $\gamma$ drops to $2$. This model is chosen because our spectral
algorithm requires the NS to have $\gamma\leq 2$ at the surface
(see more discussion in Appendix); realistic NSs do have
$\gamma\lesssim 2$ near the surface.  This model should be a
reasonably good approximation for a $\gamma=3$ polytrope, thus we call
it the ``$\gamma\simeq 3$" model. For both $\gamma=1.5$ and $\gamma\simeq3$
models we also assume a constant $\Gamma-\gamma$, so the scaling
\eqref{scale_g_1} - \eqref{scale_g_2} can still be used.

\begin{figure}[!]
\centering
\includegraphics[width=.5\textwidth]{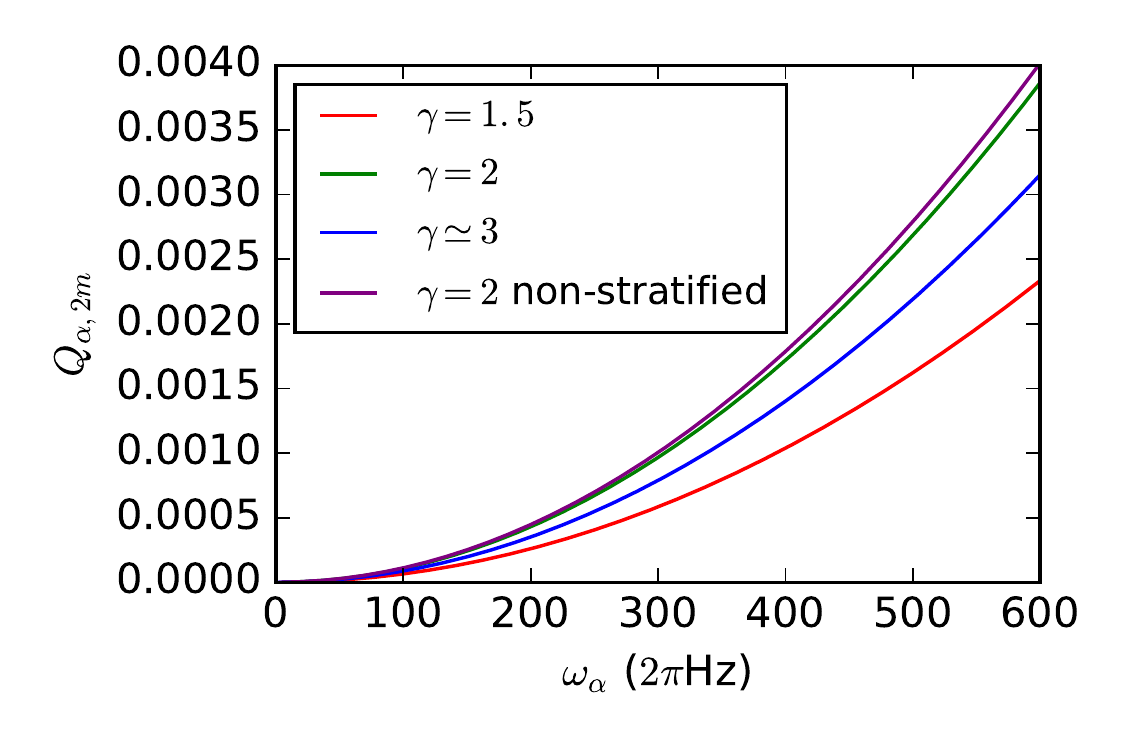}
\caption{Tidal coupling coefficient $Q_{\alpha,2m}$ of the $j=2,n=1$
  g-mode as a function of mode frequency $\omega_\alpha$ for different
  density and stratification models. In this figure we consider a
  $M=1.4M_\odot, R=10$~km NS. The ``$\gamma=2$ non-stratified" curve
  corresponds to the model with a non-stratified envelope (see Section
  IV.A.3).}
\label{g_mode_Q}
\end{figure}

Comparing the results for the three different density models, we see
that the scaled frequency and tidal coupling coefficient are both smaller for
larger $\gamma$. However, a lower mode frequency implies tidal resonance at a larger 
binary separation, leading to a larger phase shift (for the same $Q_{\alpha,2m}$);
see Eq.~(\ref{eq:orbchange}).
Figure \ref{g_mode_Q} shows the dependence of $Q_{\alpha,2m}$ on the 
mode frequency $\omega\al$ for different NS density models. We see that for the three
models studied here ($\gamma=1.5,\gamma=2$ and $\gamma\simeq 3$),
the $\gamma=2$ model yields the largest tidal coupling for given $\omega\al$.

\subsubsection{Effect of Non-Stratified Stellar Envelope}

Realistic NSs do not have a constant $\Gamma-\gamma$
throughout the star. To see how non-constant stratification affects
the g-modes, here we consider a model 
where the NS envelope has a zero 
Brunt V\"ais\"al\"a frequency ($\Gamma-\gamma=0$). Specifically, we consider 
a $\gamma=2$ polytrope with stratification given by 
\be
\Gamma(r)-\gamma =
(\Gamma_0-\gamma)\left\{{1\over 2}\tanh[20(0.8-r/R)] + {1\over 2}\right\},
\label{eq:nonstrat}\ee
where $\Gamma_0$ is a constant. This gives $\Gamma\simeq \Gamma_0$ for the inner $80\%$ of
star and $\Gamma\simeq \gamma$ for the outer $20\%$. The scaled
frequency and tidal coupling coefficient for g-modes in this NS model are
given in Table \ref{g_modes_convective}, where $\bar{\omega}\al$ and
$\bar{Q}_{\alpha,2m}$ are defined by \eqref{scale_g_1} -
\eqref{scale_g_2} except we now use $\Gamma_0-\gamma$ instead of
$\Gamma-\gamma$.

Comparing Table \ref{g_modes_convective} with Table \ref{g_modes}, we
see that this non-stratified region tends to decrease the scaled frequency and
tidal coupling of g-modes. However, as shown in Figure \ref{g_mode_Q},
$Q_{\alpha,2m}$ at given $\omega\al$ is barely affected by including a
non-stratified envelope.

\begin{table}
\centering
\caption{Scaled frequency and tidal coupling coefficient for $j=2$
  pure g-modes in a $\gamma=2$ polytropic NS with a non-stratified
  envelope.\footnote{Note that $\Gamma-\gamma$ is no longer constant in this model, and
    the definitions of $\bar{\omega}\al$ and $\bar{Q}_{\alpha,2m}$ are given by 
    Eqs.~(\ref{scale_g_1})-(\ref{scale_g_2}), with $\Gamma$ replaced by $\Gamma_0$
    (see Eq.~\ref{eq:nonstrat}).}}
\label{g_modes_convective}
\begin{ruledtabular}
  \begin{tabular}{  c  c  c }
    ~ & $\bar{\omega}\al$ $(2\pi{\rm Hz})$ & $\bar{Q}_{\alpha,2m}$ \\ \hline
    $n=1$ & $\pm 120.7$  & $1.62\times 10^{-4}$ \\ 
    $n=2$ & $\pm 77.8$ & $3.3\times 10^{-5}$ \\ 
    $n=3$ & $\pm 57.1$ & $1.0\times 10^{-5}$ \\
  \end{tabular}
  \end{ruledtabular}
\end{table}

\subsubsection{Effect of Cowling Approximation}

Many works studying NS modes apply the Cowling approximation
(e.g. \cite{KokkotasStergioulas99,Passamontietal09,YuWeinberg17a}),
which ignores $\delta\Psi$. We expect that the Cowling approximation do 
not affect the result significantly when the mode has multiple nodes in $\delta\Psi$.
But for the modes we are interested in (i.e. g-modes with small $n$, and
r-modes with small $j$ which we will discuss later), $\delta\Psi$
usually have few or no node. Therefore it is necessary to investigate
whether the Cowling approximation is accurate enough in this case. Table
\ref{g_modes_cowling} shows the frequency and tidal coupling
coefficient for $j=2$ g-modes in a $\gamma=2$ polytropic NS,
calculated with the Cowling approximation. We see that applying this
approximation barely affects the mode frequency, but the tidal
coupling $Q_{\alpha,2m}$ is overestimated by $40\%$ -
$50\%$. Therefore, Cowling approximation can only provide a crude
approximation for the tidal coupling coefficient.

\begin{table}
\centering
\caption{Scaled frequency and tidal coupling coefficient for $j=2$ pure g-modes in a $\gamma=2$ polytropic NS, using the Cowling approximation.}
\label{g_modes_cowling}
\begin{ruledtabular}
  \begin{tabular}{  c  c  c }
    ~ & $\bar{\omega}\al$ $(2\pi{\rm Hz})$ & $\bar{Q}_{\alpha,2m}$ \\ \hline
    $n=1$ & $\pm 181.5$  & $4.8\times 10^{-4}$ \\ 
    $n=2$ & $\pm 124.7$ & $1.27\times 10^{-4}$ \\ 
    $n=3$ & $\pm 95.7$ & $3.9\times 10^{-5}$ \\
  \end{tabular}
  \end{ruledtabular}
\end{table}

\subsection{Pure Inertial Modes (I-Modes)}

We now consider another limiting case, when the NS has no
stratification ($\Gamma=\gamma$). In this case all low-frequency modes
are inertial modes (i-modes), whose restoring force is the Coriolis force.

Since we assume that $\Oms$ is small [compared to the characteristic
frequency $(GM/R^3)^{1/2}$], the Lagrangian displacement of an
i-mode can be expanded in terms of $\hat\Omega_s\equiv
\Oms(R^3/GM)^{1/2}$:

\eq{
\bxi_{\alpha} = \bxi_{\alpha,0} + \hat\Omega_s\bxi_{\alpha,1} + \hat\Omega_s^2\bxi_{\alpha,2}+\cdots,
}
and $\bxi_{\alpha,i}$ is independent of $\hat\Omega_s$. Similar to
Eq. \eqref{uvw_def}, we can decompose the velocity perturbation
corresponding to $\bxi_{\alpha,i}$ into $u_{jm,i},v_{jm,i}$ and
$w_{jm,i}$. For each $i$, $u_{jm,i},v_{jm,i}$ and $w_{jm,i}$ can be nonzero 
only up to some finite $j$ \cite{LockitchFriedman99}. 

Following Ref.~\cite{LockitchFriedman99}, we label
each i-mode by $m$ and the $j=j_0$ value, the latter given by the largest number such that
at least one of $u_{jm,0},v_{jm,0}$ and $w_{jm,0}$ is nonzero (this $j_0$ value is
the same as $l_0$ in \cite{LockitchFriedman99}). In general, there are
$(j_0-m+1)$ modes for each pair of $j_0$ and $m$. The value $j_0$ is also related to
the parity of the mode. Even $j_0$ corresponds to odd parity modes and
odd $j_0$ even parity modes\footnote{Here even (odd) parity means that
the scalar perturbations of the mode, such as the density
perturbation, is conserved (changes sign) upon a parity
transformation.}. 
Since only the even parity modes couple to the tidal
potential, we will only consider modes with odd $j_0$ 
(which will be simply labeled $j$ in the discussion below).

\subsubsection{Inertial Mode Scaling Relations}

Table \ref{i_modes} shows the frequency and scaled coupling
coefficient for $j=1$ and $3$ pure inertial modes for the $\gamma=1.5,~2$
and $\gamma\simeq 3$ density models. Our results show that to lowest
order of $\hat\Omega_s$, the mode frequencies $\omega\al,\sigma\al \propto \Oms$ and
\eq{
Q_{\alpha,2m}\propto \hat\Omega_s^2,
}
therefore we define the scaled tidal coupling coefficient by
\eq{
\hat Q_{\alpha,2m} = Q_{\alpha,2m} \hat\Omega_s^{-2}.\label{hatQ_def}
}
Note that $\hat Q_{\alpha,2m}$ is also independent of $M$ and $R$. The
mode frequencies agree with the result from \cite{LockitchFriedman99}
very well, and the scalings agree with the result of \cite{LaiWu06}.

Form Table \ref{i_modes} we can see that $Q_{\alpha,2m}$ is mainly
determined by $j$, and decreases as $j$ increases; the $j=3$ modes have
$Q_{\alpha,2m}$ about an order of magnitude smaller than the $j=1$
mode. Moreover, the $j\geq 5$ modes have $Q_{\alpha,2m}$ about one order
of magnitude smaller than the $j\leq 3$ modes. Therefore, the most important 
contribution to the GW phase shift comes from $j\leq 3$ i-modes.

The $j=m$ i-modes are often referred to as Rossby modes or r-modes; we will call them r-modes
from now on to distinguish them from other i-modes.
The $j=m=1$ r-mode is of particular interest. 
It has $\omega_\alpha/\Omega_s=-1$ and $\sigma_\alpha/\Omega_s=0$ to the lowest order of
$\Oms$, and this result is independent of the NS density profile. 
Our numerical calculation shows that this mode has
\eq{
\sigma\al \propto \hat\Omega_s^2\Oms,
}
and the proportionality constant can be found in Table
\ref{i_modes}. This coefficient should also be affected by the
centrifugal distortion of the NS due to rotation, which we do not
include in our calculation for Table \ref{i_modes}. Kokkotas \&
Stergioulas \cite{KokkotasStergioulas99} gave an analytical
calculation of the frequencies of r-modes for an uniform density star
with the effect of centrifugal distortion included (but they used
Cowling approximation); they found that for this $j=m=1$ r-mode
\eq{
\sigma\al = -\frac{3}{4}\hat\Omega_s^2\Oms.
}
The scaling of this result is the same as ours, but the coefficient is
very different. Importantly, our calculation shows that this mode is
prograde in the inertial frame ($\sigma\al>0$ for all our NS models) while theirs retrograde
($\sigma\al<0$). This suggests that including centrifugal distortion
may nontrivially affect the result. We will discuss the effect of
distortion by re-calculating this mode in the following subsection.

\begin{table*}
\centering
\caption{Scaled frequency and tidal coupling coefficient for $j=1$ and
  $3$ pure i-modes in NSs with different density models.\footnote{This
    calculation ignores centrifugal distortion (but includes the gravitational perturbation),
    which is a good approximation for all modes except for the $j=m=1$ mode. See Section IV.B.1-2
    for discussion.}}
\label{i_modes}
\begin{ruledtabular}
\begin{tabular}{c@{\hskip .5in}ccc@{\hskip .5in}ccc@{\hskip .5in}ccc}
~&\multicolumn{3}{c@{\hskip .5in}}{$\gamma=1.5$} & \multicolumn{3}{c@{\hskip .5in}}{$\gamma=2$} & \multicolumn{3}{c}{$\gamma\simeq3$\footnote{Same as the $\gamma\simeq 3$ model used in Table \ref{g_modes}.}} \\
~ & $\omega\al/\Oms$ & $\sigma\al/\Oms$ & $\hat{Q}_{\alpha,2m}$\footnote{$\hat{Q}_{\alpha,2m}=Q_{\alpha,2m} \hat\Omega_s^{-2}$ as defined in Eq.~\eqref{hatQ_def}.} 
& $\omega\al/\Oms$ & $\sigma\al/\Oms$ & $\hat{Q}_{\alpha,2m}$ 
& $\omega\al/\Oms$ & $\sigma\al/\Oms$ & $\hat{Q}_{\alpha,2m}$\\
$j=1,m=1$&-1.000&0.32$\hat\Omega_s^2$&0.137	&-1.000&0.66$\hat\Omega_s^2$&0.370 	&-1.000&0.90$\hat\Omega_s^2$&0.559 \\
$j=3,m=2$&0.827&2.827&0.031 		&0.556&2.556&0.015 		&0.405&2.405&0.005 \\
~		&-1.034&0.966&0.010 		&-1.100&0.900&0.010 		&-1.151&0.849&0.005 \\
$j=3,m=1$&1.184&2.184&0.026 		&1.032&2.032&0.017 		&0.938&1.938&0.007 \\
~		&-0.746&0.254&0.014 		&-0.690&0.310&0.009 		&-0.656&0.344&0.004 \\
~		&-1.545&-0.545&0.020 		&-1.613&-0.613&0.014 		&-1.655&-0.655&0.006
\end{tabular}
\end{ruledtabular}
\end{table*}

\subsubsection{Effect of Centrifugal Distortion}

For all of our calculations (except for this subsection), we assume that the
NS is spherical and ignore distortion due to the finite rotation
rate. Here we calculate the modes in a distorted NS to study if
including this distortion can nontrivially affect the result. As discussed above,
this may be especially important for the $j=m=1$ r-mode. For the
stellar density profile, we assume that the NS is barotropic and obtain the
distorted profile iteratively, similar to the method in
\cite{Hachisu86}. The method of calculating modes in a distorted star
is summarized in \cite{Reeseetal06}.

We find that the correction to $\omega\al/\Oms$ and $\hat Q_{\alpha,2m}$ 
due to distortion are both of order $\hat\Omega_s^2$. As a result, this correction 
is unimportant in most cases. The only exception is the $j=m=1$ r-mode, where
$\sigma\al/\Oms$ is also of order $\hat\Omega_s^2$. Figure
\ref{r_distortion} shows the inertial-frame frequency $\sigma\al$ for
two different density models when different approximations are
used. We see that when we ignore distortion, we get $\sigma\al>0$ and
$\sigma\al\propto \hat\Omega_s^2\Oms$. When distortion is included,
however, $\sigma\al$ becomes very close to 0: Our results show that
$|\sigma\al|/\Oms\lesssim 5\times 10^{-5}$, which is comparable to the
estimated numerical error due to the finite grid size of our calculation. 
This suggests that this $j=m=1$ r-mode likely has $\sigma\al/\Oms=0$ or $\mathcal
O(\hat\Omega_s^4)$. Moreover, this result is independent of the
density profile used: In Figure \ref{r_distortion} we use two
different density profiles ($\gamma=2$ and $\gamma\simeq 3$), and
$\sigma\al$ is clsoe to zero for both models.

\subsubsection{Effect of Cowling Approximation}

\begin{figure}[!]
\centering
\includegraphics[width=.5\textwidth]{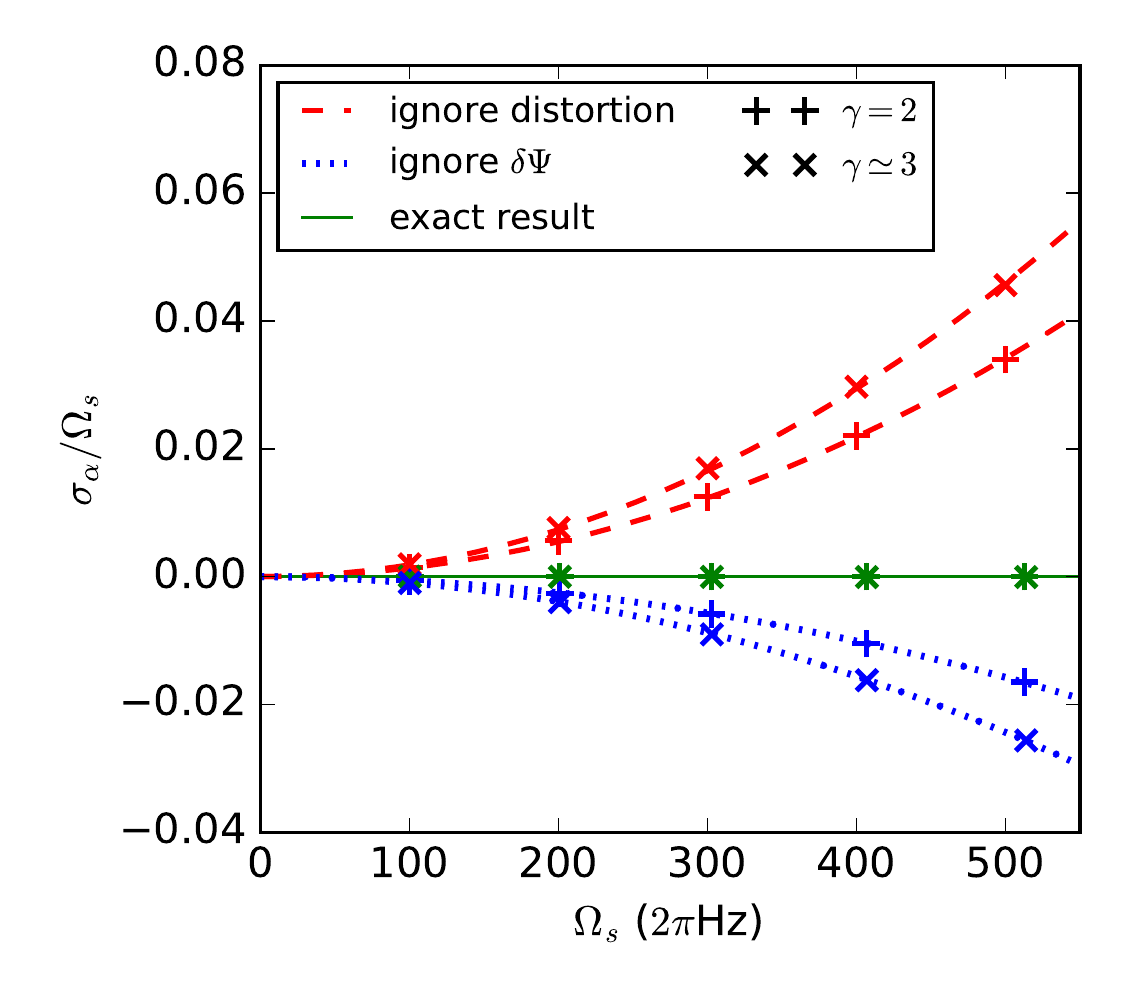}
\caption{The inertial-frame frequency $\sigma_\alpha$ (in units of the spin frequency $\Omega_s$)
  for the $j=m=1$ r-mode in a $M=1.4M_\odot$ and $R=10$km NS, calculated using different
  approximations. Satires and crosses mark the $\gamma=2$ and
  $\gamma\simeq 3$ density models, respectively. The red dashed curves
  correspond to results for a spherical star (i.e. the rotational
  distortion is ignored) and blue dotted curves correspond to results
  with Cowling approximation (i.e. $\delta\Psi$ is ignored). The
  curves are best-fitting results assuming $\sigma\al/\Oms\propto
  \hat\Omega_s^2$; we see that this fits the numerical data points well. The
  green lines show the ``exact'' results where both $\delta\Psi$ and distortion
  are included: These results have $\sigma\al/\Oms\simeq
  0$; the deviation from zero is comparable to the numerical error 
  ($|\sigma\al|/\Oms\lesssim 5\times10^{-5}$ for
  $\Oms<500\cdot 2\pi$Hz), so it is likely that this mode has exactly
  zero inertial-frame frequency or $\sigma\al/\Oms$ is of order
  $\hat\Omega_s^4$ or higher.}
\label{r_distortion}
\end{figure}

Kokkotas \& Stergioulas \cite{KokkotasStergioulas99} calculated the
inertial-frame frequencies $\sigma\al$ of $j=m$ r-modes in a rotationally
distorted, incompressible (uniform density) star under Cowling approximation
(i.e. ignore $\delta\Psi$). To compare with their result, we also
calculate $\sigma\al$ using Cowling approximation in a distorted
$\gamma=2$ or $\gamma\simeq 3$ NS model (see Fig.~\ref{r_distortion}). We see
that under Cowling approximation, the $j=m=1$ r-mode has $\sigma\al/\Oms\propto
\hat\Omega_s^2$ and the mode is retrograde ($\sigma\al<0$), which
has the same sign as the result (for incompressible stars) in
\cite{KokkotasStergioulas99}. For our NS models with $\gamma=1.5,~2$
and $\gamma\simeq 3$, this mode has $\sigma\al/(\hat\Omega_s^2\Oms) =
-0.11,-0.30$ and $-0.46$ respectively, whereas for $\gamma=\infty$, the 
analytic (Cowling approximation) result is $\sigma\al/(\hat\Omega_s^2\Oms) =-0.75$.

Similar to the case for g-modes, Cowling approximation barely affect
the result of $\omega\al$ and $\sigma\al$ up to $\mathcal O(\Oms)$,
but produces a nontrivial error in $Q_{\alpha,2m}$ (relative error
$\gtrsim 30\%$ for most modes in Table \ref{i_modes}). Therefore,
Cowling approximation should not be used if we need to obtain a
relatively accurate result for tidal coupling and GW phase shift.

\subsubsection{Discussion on the $j=m=1$ R-Mode}

As noted above (see Fig.~\ref{r_distortion}), in our ``exact'' calculation, which 
takes account of the rotational distortion and gravitational potential perturbation,
the $j=m=1$ r-mode has zero frequency in the inertial frame. Calculations that neglect
either of these effects would give an incorrect result ($\sigma_\alpha/\Omega_s
\sim \hat\Omega_s^2$). This is important because such incorrect result would imply
that for a rapidly rotating NS the mode frequency would lie in the LIGO sensitivity band 
and the resonance could be detectable due to the large GW phase shift it inflicts \cite{LaiWu06}.

It is useful to understand why this $j=m=1$ mode has zero frequency.
In the $\Omega_s\rightarrow 0$ limit, the velocity perturbation
associated with this mode is (see Eq.~\ref{uvw_def})
\be
\delta\bv \propto r \left({1\over\sin\theta}{\partial Y_{11}\over\partial\phi}\mathbf{e}_\theta 
- {\partial Y_{11}\over\partial\theta}\mathbf{e}_\phi\right).
\ee
This corresponds to a ``spin-over'' perturbation, i.e., the
equilibrium rotation around a different axis.  With a finite rotation
rate ($\Omega_s\neq 0$), there also exists such a ``spin-over''
mode. When all the effects (distortion and $\Delta\Psi$) are included,
the perturbed state should have the same energy as the unperturbed
state, and thus the mode should have zero inertial-frame frequency.

\subsection{Mixed Modes (Inertial-Gravity Modes)}

\begin{figure}[!]
\centering
\includegraphics[width=.5\textwidth]{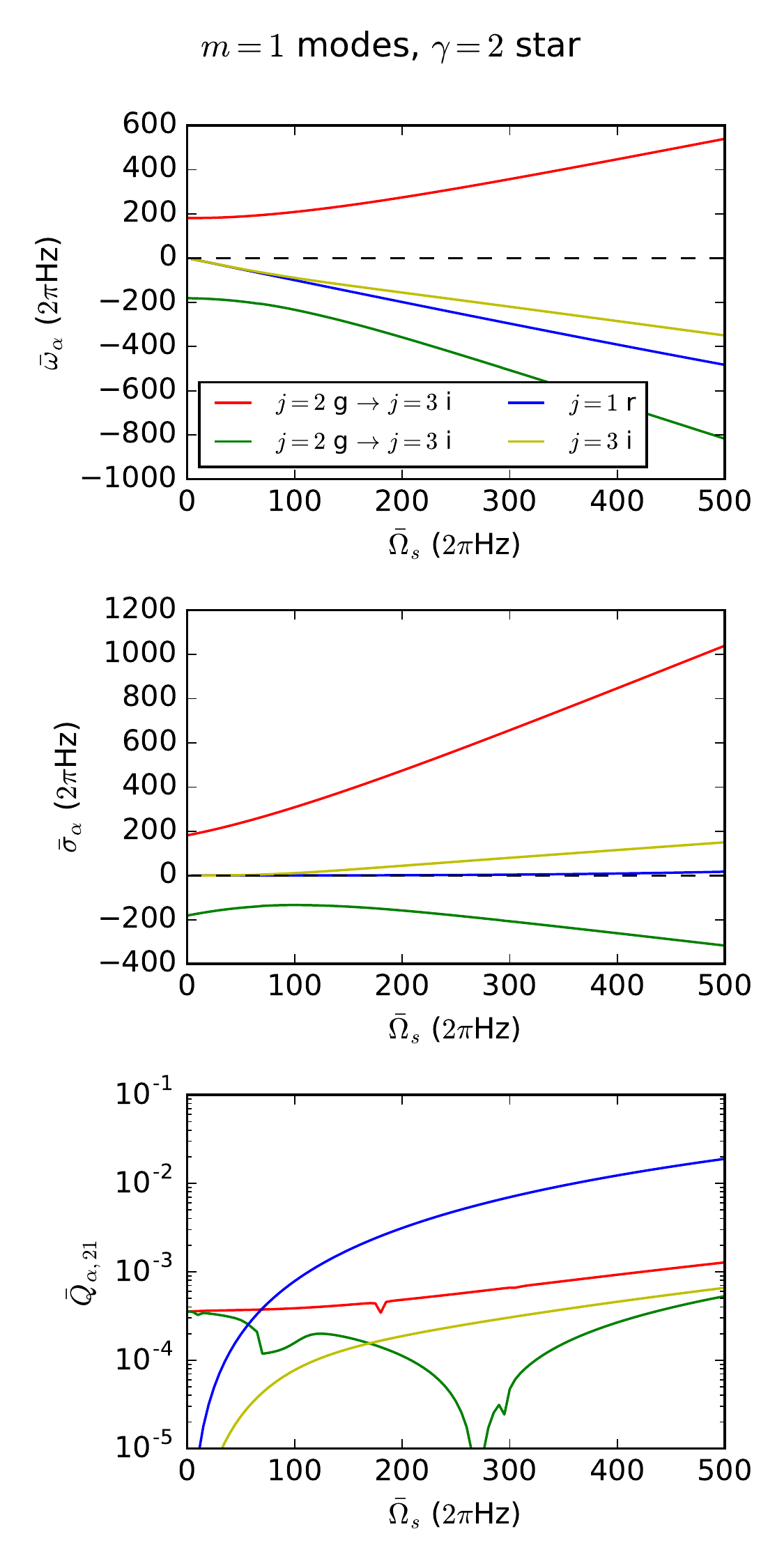}
\caption{Scaled rotating (inertial) frame frequency $\bar\omega_s$
  ($\bar\sigma_s$) and tidal coupling coefficient $\bar Q_{\alpha,21}$
  for the $m=1$ mixed modes in a rotating and stratified $\gamma=2$
  polytropic NS. The results in this figure are exact for
  $\Gamma-\gamma=0.01$, and are accurate to the lowest order
  [$\hat\Omega_s$ and $(\Gamma-\gamma)^{1/2}$ order for frequencies,
    and $\hat\Omega_s^2$ and $(\Gamma-\gamma)$ order for
    $Q_{\alpha,21}$] otherwise. The mode is labeled by the pure g-mode
  and pure i-mode they are similar to in the small $\Oms$ and large $\Oms$
  limits; e.g. ``$j=2$ g $\to$ $j=3$ i" denotes a mode that is similar
  to a $j=2$ g-mode when $\Oms\to 0$ and is similar to a $j=3$ i-mode
  when $\Oms$ is large. Modes that have zero frequency at $\Oms=0$ are
  labeled by the i-mode they are similar to at large $\Oms$;
  e.g. ``$j=3$ i" denotes a mode that becomes similar to a $j=3$
  i-mode at large $\Oms$.}
\label{m1_mix}
\end{figure}

\begin{figure}[!]
\centering
\includegraphics[width=.5\textwidth]{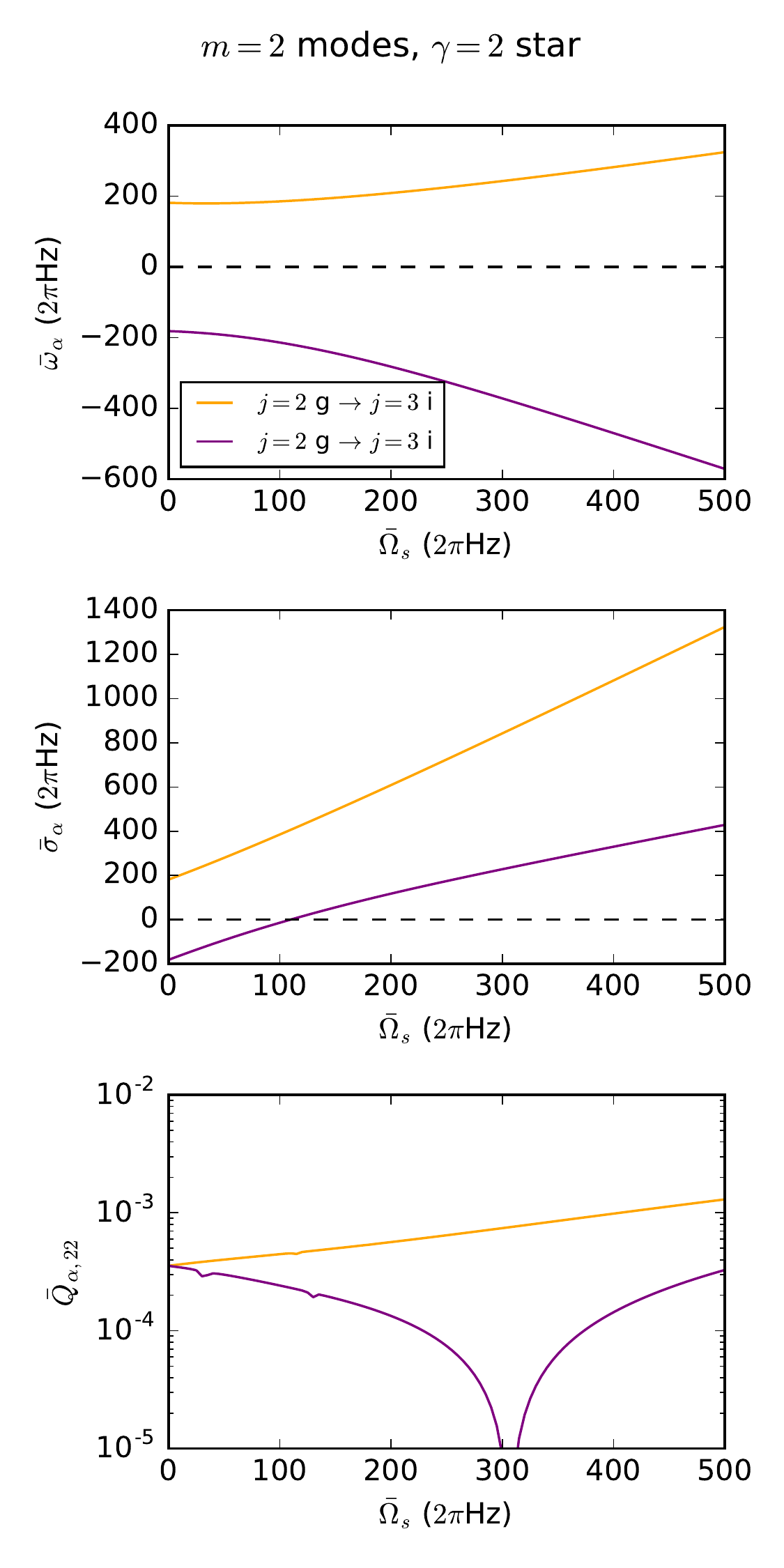}
\caption{Same as Figure \ref{m1_mix} but for the $m=2$ modes.}
\label{m2_mix}
\end{figure}

We now consider modes in rotating, stratified NSs. In such stars,
low-frequency modes involve mixing between g-modes and i-modes. Most of
these mixed modes become similar to i-modes when stratification is
weak or $\Oms$ is large, and similar to g-modes when 
stratification is strong or $\Oms$ is small.
Figure \ref{m1_mix} shows our numerical results
for $\omega_\alpha,~\sigma_\alpha$ and $Q_{\alpha,21}$ for the $m=1$
mixed modes with strongest tidal coupling. As in Section III, we use the scaled rotation
rate, mode frequencies and tidal coupling coefficient given by
\ea{
&{\bar\omega\al\over\omega\al} = {\bar\sigma\al\over\sigma\al} = 
{\bar\Omega_s\over\Oms} = \left(\frac{\Gamma-\gamma}{0.01}\right)^{-1/2}M_{1.4}^{-1/2}R_{10}^{3/2},\label{eq:scaleomega}\\
&{\bar Q_{\alpha,2m}\over Q_{\alpha,2m}} = \left(\frac{\Gamma-\gamma}{0.01}\right)^{-1}.
\label{eq:scaleQ}
}
The scaled quantities are independent of $\Gamma-\gamma$, $M$ and $R$
to the lowest order. This allows our results to be easily adapted for
different values of $\Gamma-\gamma$, $M$ and $R$.

Four modes shown in Figure \ref{m1_mix}. Two of these are the $j=2,m=1,n=1$ g-modes 
in the limit $\Oms\to 0$. As $\Oms$ increases, the frequencies, tidal coupling 
coefficients and perturbation profiles (eigenfunctions) of the modes change characters
and become mixed with i-modes. Towards the right side of the plot (high $\Omega_s$), 
the effect of rotation dominates and these two modes asymptotically become two of the 
three $j=3,m=1$ i-modes discussed in Section B. 
The other two modes shown in Fig.~\ref{m1_mix} are another $j=3,m=1$ i-mode and the 
$j=m=1$ r-mode. As $\Oms\to 0$, they both have $\omega\al\to 0$, indicating that they do not 
become g-modes in this limit. 

\begin{figure}[!]
\centering
\includegraphics[width=.5\textwidth]{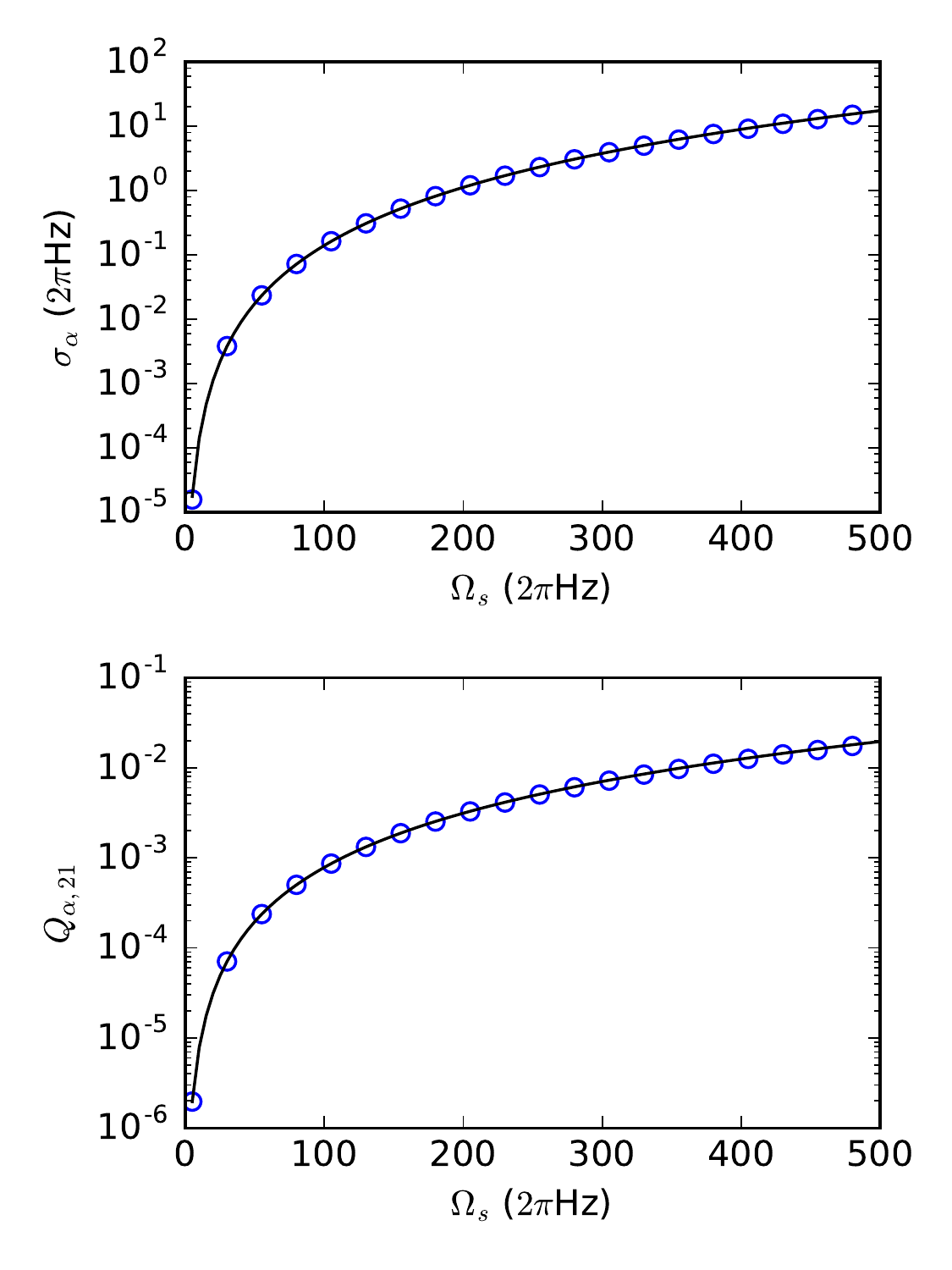}
\caption{Inertial-frame frequency $\sigma\al$ and tidal coupling
  coefficient $Q_{\alpha,21}$ of the $j=m=1$ r-mode in a $\gamma=2$
  polytropic NS with $M=1.4M_\odot$ and $R=10$km. Blue circles are
  the results for a stratified NS with $\Gamma-\gamma=0.01$ (corresponding
  to the blue curve in Figure \ref{m1_mix}), and black curves are
  the results for an unstratified NS. We see that the two results are
  identical, even at $\Oms\to 0$ where the effect of stratification is
  much stronger than that of rotation.
  Note that the results shown in this figure (and in other figures of Section IV.C)
  do not include the effect of the rotational distortion; when the effect is included,
  $\sigma_\alpha$ for the $j=m=1$ r-mode is zero to the accuracy of our calculation.
}
\label{mixed_r}
\end{figure}

Figure \ref{mixed_r} shows the comparison of $\sigma\al$ and
$Q_{\alpha,21}$ of the $j=m=1$ r-mode in a stratified NS and an
unstratified NS; we see that there is no noticeable difference between
the two results, suggesting that this mode is unaffected by
stratification. Since we already know (see Section IV.B) that this
mode has an inertial-frame frequency that satisfies $|\sigma\al|\ll
\hat\Omega_s^2\Oms$ when there is no stratification, we conclude that
the same inequality applies for realistic stratified NSs.

\begin{figure}[!]
\centering
\includegraphics[width=.5\textwidth]{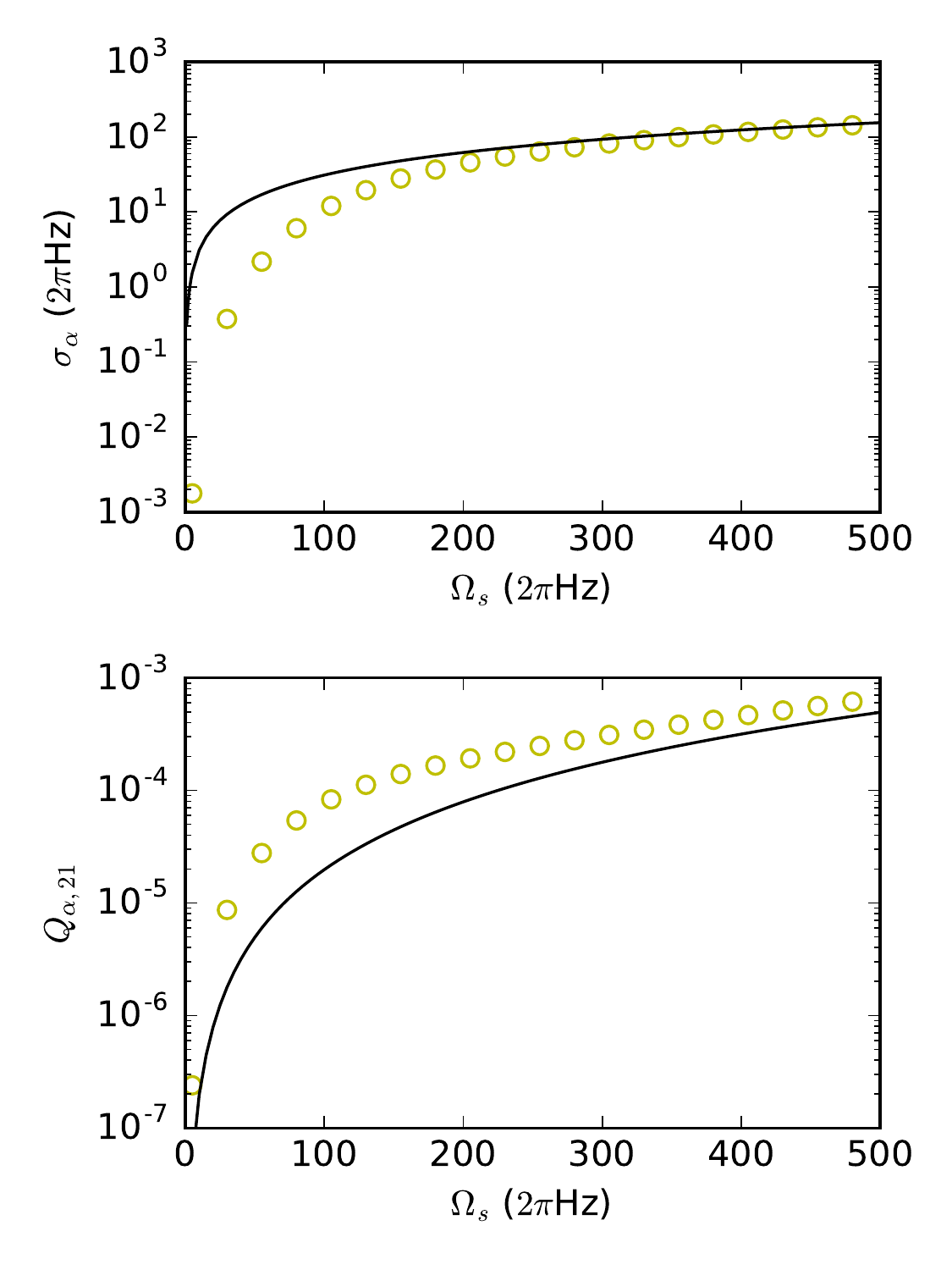}
\caption{Inertial-frame frequency $\sigma\al$ and tidal coupling
  coefficient $Q_{\alpha,21}$ of the $j=3,m=1$ i-mode that does not
  become a g-mode in a stratified NS when $\Oms=0$ (the yellow curve in
  Fig.~\ref{m1_mix}) in a $\gamma=2$ polytropic NS with
  $M=1.4M_\odot$ and $R=10$km. 
The yellow circles are the results for a
  stratified NS with $\Gamma-\gamma=0.01$ (corresponding to the yellow
  curve in Fig.~\ref{m1_mix}), and the black curves are results for an
  unstratified NS. We see that $\sigma\al$ is significantly decreased
  for small $\Oms$ and the tidal coupling coefficient is 
  increased due to the stratification.}
\label{mixed_i}
\end{figure}

The $j=3,~m=1$ i-mode is affected by stratification in an interesting
way.  Figure \ref{mixed_i} shows the comparison of $\sigma\al$ and
$Q_{\alpha,21}$ of this $j=3, m=1$ i-mode in a stratified NS and an
unstratified NS.  We see that stratification causes $\sigma\al$ to
decrease and $Q_{\alpha,21}$ to increase.  The effect of
stratification is most significant when $\Oms\to 0$: the decreased
$\sigma\al$ and increased $Q_{\alpha,21}$ makes the GW phase shift
converge to a finite value as $\Oms\to 0$, while when there is no
stratification the GW phase shift goes to zero at $\Oms\to 0$. We will
discuss the GW phase shift of this mode in more detail in Section V.

Figure \ref{m2_mix} is similar to Fig.~\ref{m1_mix} but shows the $m=2$
mixed modes. It shows two modes which are the two $j=2,~m=2,~n=1$
g-modes for $\Oms\to 0$. Similar to the $m=1$ g-modes, their frequencies
and perturbation profiles change as $\Oms$ increases, and for large
$\Oms$ they asymptotically become the two $j=3,~m=2$ i-modes. Note that
the retrograde g-mode becomes a prograde i-mode as $\Oms$ increases (the purple lines in
the figure), therefore the inertial frame frequency $\sigma\al$ crosses zero at
some intermediate $\Oms$ (for our NS model this happens at
$\bar\Omega_s\simeq 110\cdot 2\pi$Hz). When $\sigma\al$ is small, the
GW phase shift can be significantly boosted as we see from
Eq.~\eqref{eq:orbchange} that $\Delta\Phi\propto \sigma\al^{-1}$.

A major goal of this paper is to study whether tidal coupling can be
increased by the mixing between g-modes and i-modes in the regime when
the effects of stratification and rotation are comparable. In this
regime the perturbation profiles of the modes can be significantly
different from pure g-modes or pure i-modes, which in principle may
allow the corresponding $Q_{\alpha,2m}$ to increase. However, Figures
\ref{m1_mix} and \ref{m2_mix} show that the tidal coupling coefficient
does not significantly increase in this regime; instead, for some
modes the tidal coupling is suppressed (e.g. the green curve in Figure
\ref{m1_mix} and the blue curve in Figure \ref{m2_mix}). Therefore,
the mixing between g-modes and i-modes in general does not increase
the GW phase shift by increasing the tidal coupling coefficient, but
mainly by changing the inertial-frame mode frequency.

\begin{figure}[!]
\centering
\includegraphics[width=.5\textwidth]{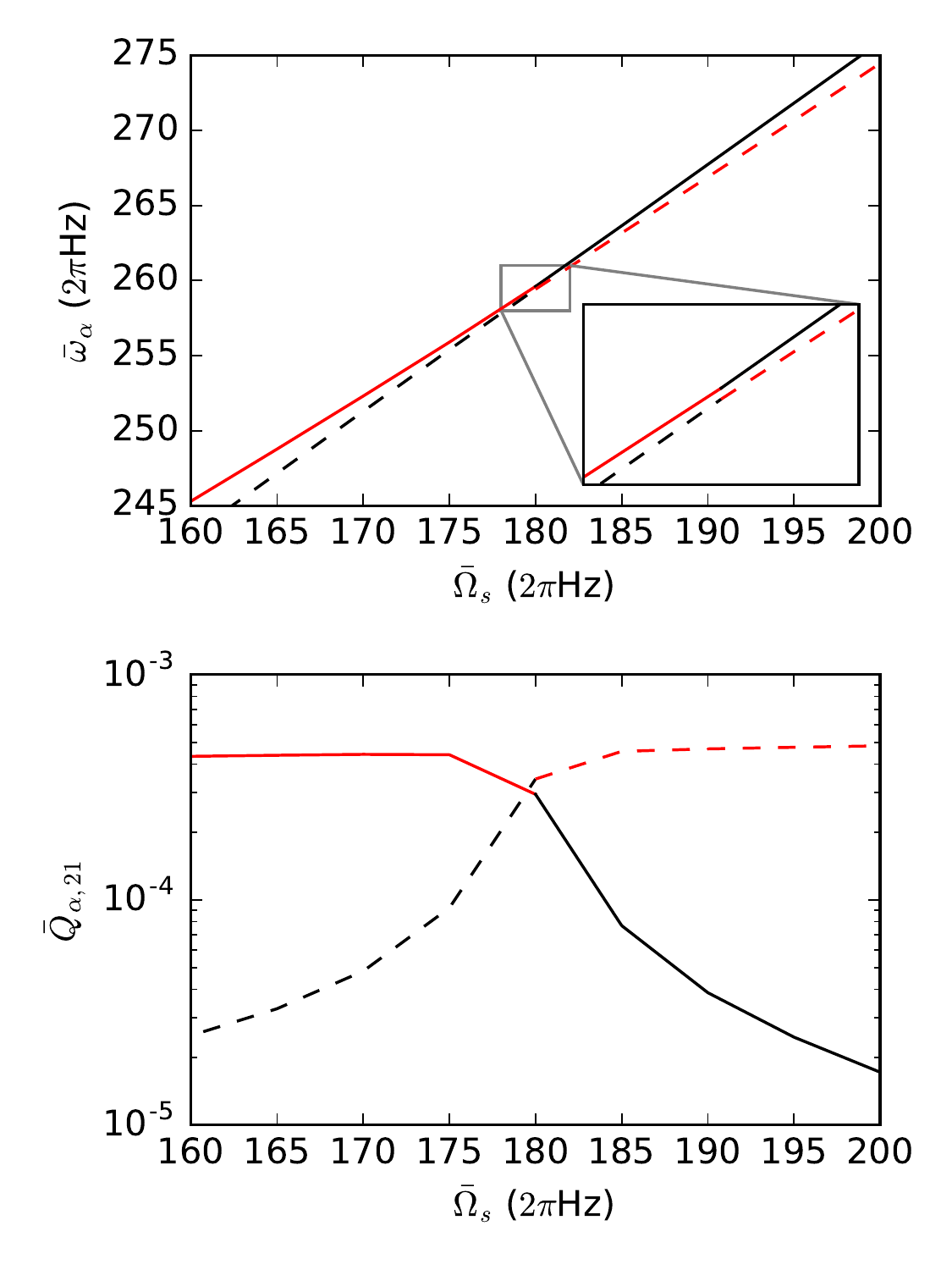}
\caption{``Crossing" of two adjacent modes. The solid and dashed curves
  mark two adjacent modes. We see that as they become close to each
  other at $\simeq 180\cdot 2\pi$Hz, they avoid crossing each other
  but exchanges perturbation profiles. The exchange is apparent if we
  look at $\bar Q_{\alpha,21}$ of the two modes. In order to trace the
  mode we are most interested in (i.e., 
the mode with larger tidal coupling), we choose to consider the red colored
  branches as a single mode. The mode shown here is the $j=2,~m=1$ modified
  g-mode with positive frequency shown in Fig.~\ref{m1_mix} (the red
  curve).}
\label{crossing}
\end{figure}

We note that as a result of mode mixing, there are some ambiguities in 
tracing the evolution of different modes as $\Oms$ varies. 
In fact, in Figure 
\ref{m1_mix} and Figure \ref{m2_mix} some of the curves do not strictly follow 
a single mode. As the example depicted in Fig.~\ref{crossing} shows, two
adjacent modes may exchange their perturbation profiles
and the tidal coupling coefficients due to mixing. When this happens, we cross
into the other mode which preserves the perturbation profile we are interested in
and has a larger tidal coupling. This is the reason for the few small dips 
in bottom panels of Figs.~\ref{m1_mix} and \ref{m2_mix}.

\section{GW phase shift due to tidal resonance}

Given the frequency and tidal coupling coefficient of the NS oscillation 
modes, we can calculate the GW phase shift $\Delta\Phi$ due to each tidal resonance. 
In this section we first consider $\Delta\Phi$ due to pure (non-rotating)
g-mode resonance in different NS models. Then we 
discuss $\Delta\Phi$ for mixed modes, emphasizing two cases
where the mixing between g-modes and i-modes affects the GW phase shift
in a nontrivial way.

\begin{figure}[!]
\centering
\includegraphics[width=.5\textwidth]{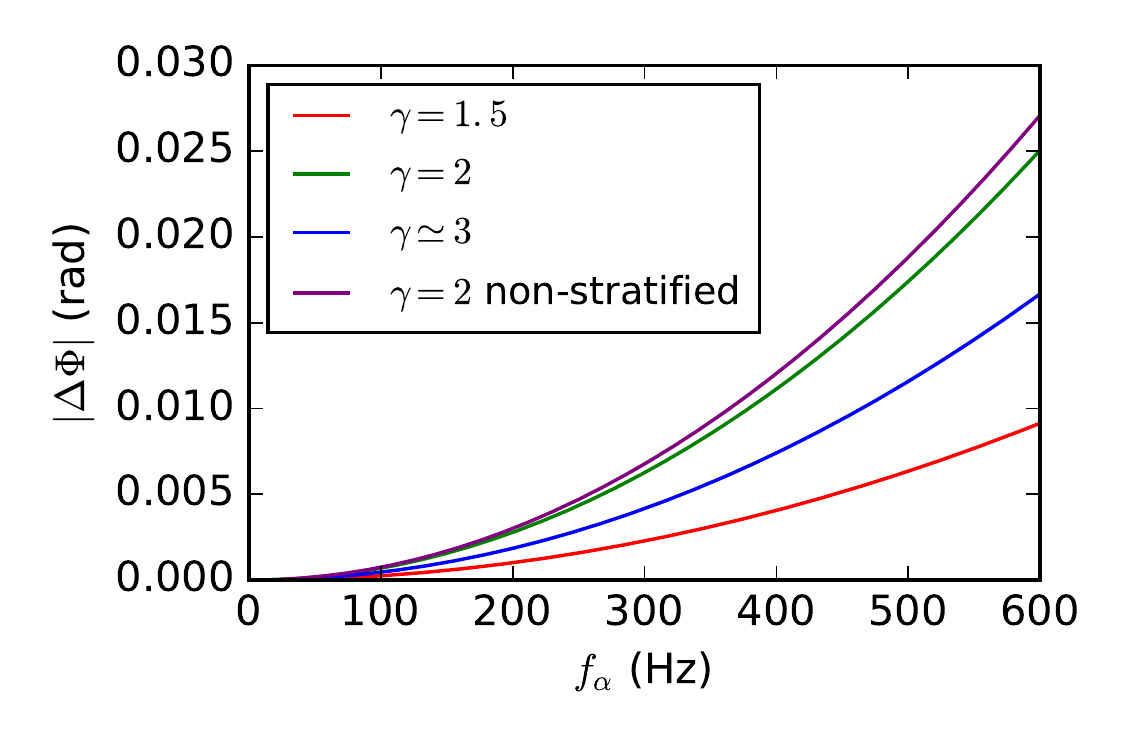}
\caption{GW phase shift $|\Delta\Phi|$ due to tidal
  resonance with the $j=2,n=1$ g-mode as a function of mode frequency
  $f_\alpha = \omega_\alpha/2\pi$ (also the GW frequency) for different NS density and
  stratification models. The sign of $\Delta\Phi$ is negative.  In
  this figure we consider a $M=1.4M_\odot,~R=10$km NS in an equal-mass
  binary with aligned spin ($\Theta=0$). The ``$\gamma=2$ non-stratified" curve corresponds to the
  model with a non-stratified envelope (see Section IV.A.3).}
\label{g_mode_Phi}
\end{figure}

\subsection{Pure g-Modes}

We begin by considering the $\Oms=0$ case, and all
low-frequency modes are pure g-modes. 
From Eqs.~(\ref{eq:orbchange}), (\ref{scale_g_1}) and (\ref{scale_g_2}), we find that for aligned spin ($\Theta=0$),
\begin{eqnarray}
&&\Delta\Phi=-0.060\left({R_{10}\over M_{1.4}}\right)^{\!5}{2\over q(1+q)}
\left({\Gamma-\gamma\over 0.01}\right)\nonumber\\
&&\qquad\quad\times \left(\!{{\bar f}_\alpha\over 100\,{\rm Hz}}\!\right)^{\!\!-2}
\left({{\bar Q}_{\alpha,22}\over 10^{-3}}\right)^{\!2}\nonumber\\
&&\qquad = -0.060\left({R_{10}^8\over M_{1.4}^6}\right){2\over q(1+q)}
\left({f_\alpha\over 100\,{\rm Hz}}\right)^2\nonumber\\
&&\qquad\quad\times \left(\!{{\bar f}_\alpha\over 100\,{\rm Hz}}\!\right)^{\!\!-4}
\left({{\bar Q}_{\alpha,22}\over 10^{-3}}\right)^{\!2},
\label{eq:gmode-phase}\end{eqnarray}
where $f_\alpha=\omega_\alpha/(2\pi)$ is the mode frequency, 
and ${\bar f}_\alpha={\bar\omega_\alpha}/(2\pi)$ and ${\bar Q}_{\alpha,22}$ can be
directly read off from Tables I-III for different NS models.
Figure \ref{g_mode_Phi} shows
the magnitude of the phase shift $\Delta\Phi$ due to tidal resonance with the
$j=2,m=2,n=1$ g-mode in a canonical NS binary ($M=1.4M_\odot$, $R=10$~km and $q=1$)
with different density and stratification profiles, as a function of the mode
frequency $f_\alpha$ (which is also the inertial frame frequency, and
is equal to the GW frequency). We see that among the three density
models we considered, $\gamma=2$ gives the largest GW phase shift for
given $f_\alpha$. Meanwhile, including a non-stratified envelope barely
affects the relation between the GW phase shift and $f_\alpha$. For
$f_\alpha$ at $\sim 300$~Hz, we find $|\Delta\Phi|
\lesssim 0.01$ for the canonical NS binaries.

It is important to recognize the strong dependence of Eq.~(\ref{eq:gmode-phase})
on the mass and radius of the NS. For example, if we consider a 
$M=1.2M_\odot$, $R=13$~km NS (which is entirely allowed or even preferred
by empirically constrained nuclear equations of state; see \cite{Steineretal13}; 
also note that the measured NS mass ranges from $1.17M_\odot$ to $2M_\odot$),
the phase shift in Fig.8 should be increases by a factor ($\propto R^8/M^6$) 
of 20.6 !

\begin{figure}[!]
\centering
\includegraphics[width=.5\textwidth]{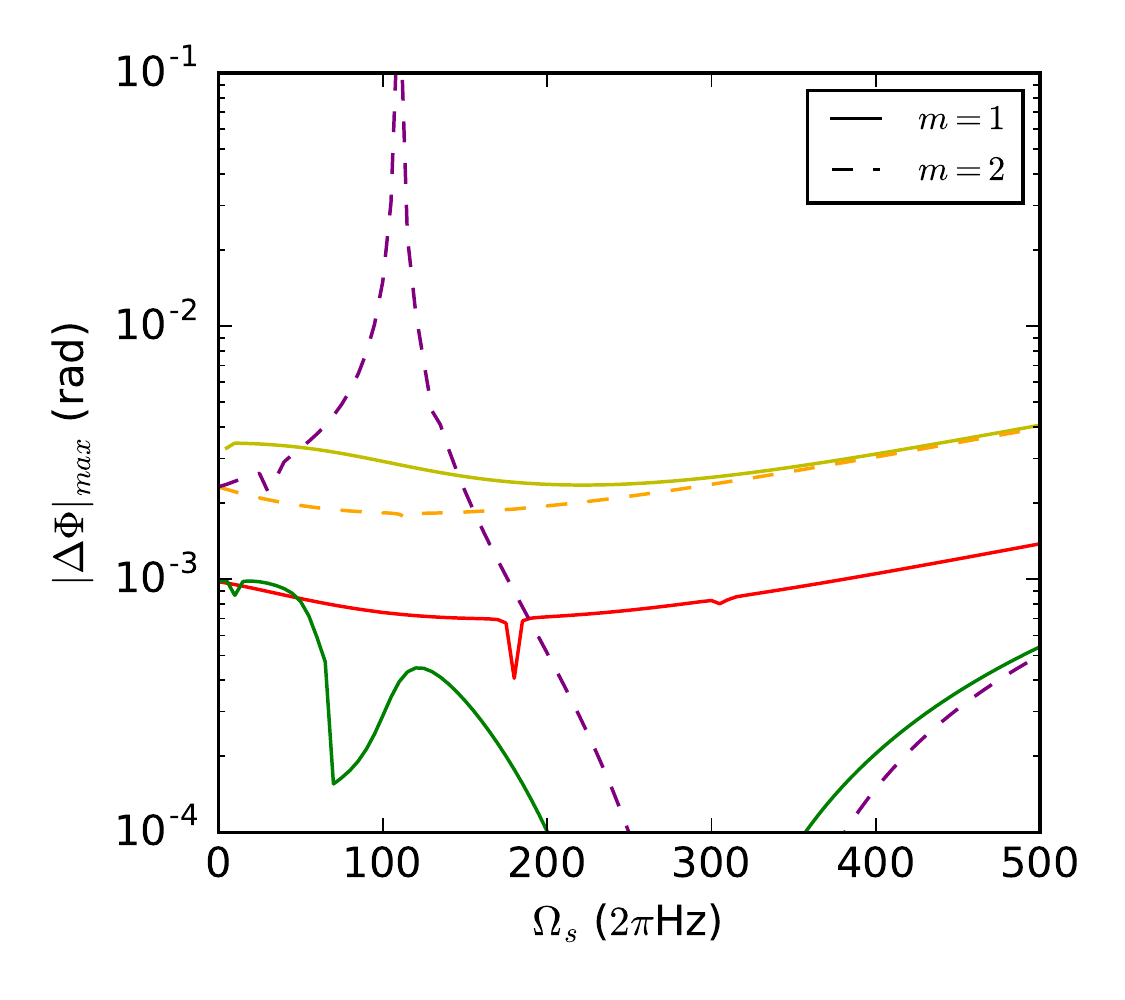}
\caption{Maximum GW phase shift $|\Delta\Phi|_{\rm max}$
  (i.e. assuming a spin-orbit inclination which gives the largest
  $|\Delta\Phi|$) as a function of the NS rotation rate $\Oms$, for
  different modes in a $\gamma=2$ polytropic NS with
  $M_{1.4}=R_{10}=1$ and $\Gamma-\gamma=0.01$, with an equal-mass
  companion. The solid (dashed) lines denote the $m=1$ ($m=2$) modes.
  Each line corresponds to the mode with the same color shown in
  Fig.~\ref{m1_mix} (for $m=1$) or Fig.~\ref{m2_mix} (for $m=2$). Note
  that the $j=m=1$ r-mode (the blue lines in Fig.~\ref{m1_mix}) is not
  included here because its inertial-frame frequency is essentially
  zero.
}
\label{Phi_mixed}
\end{figure}

\begin{figure}[!]
\centering
\includegraphics[width=.5\textwidth]{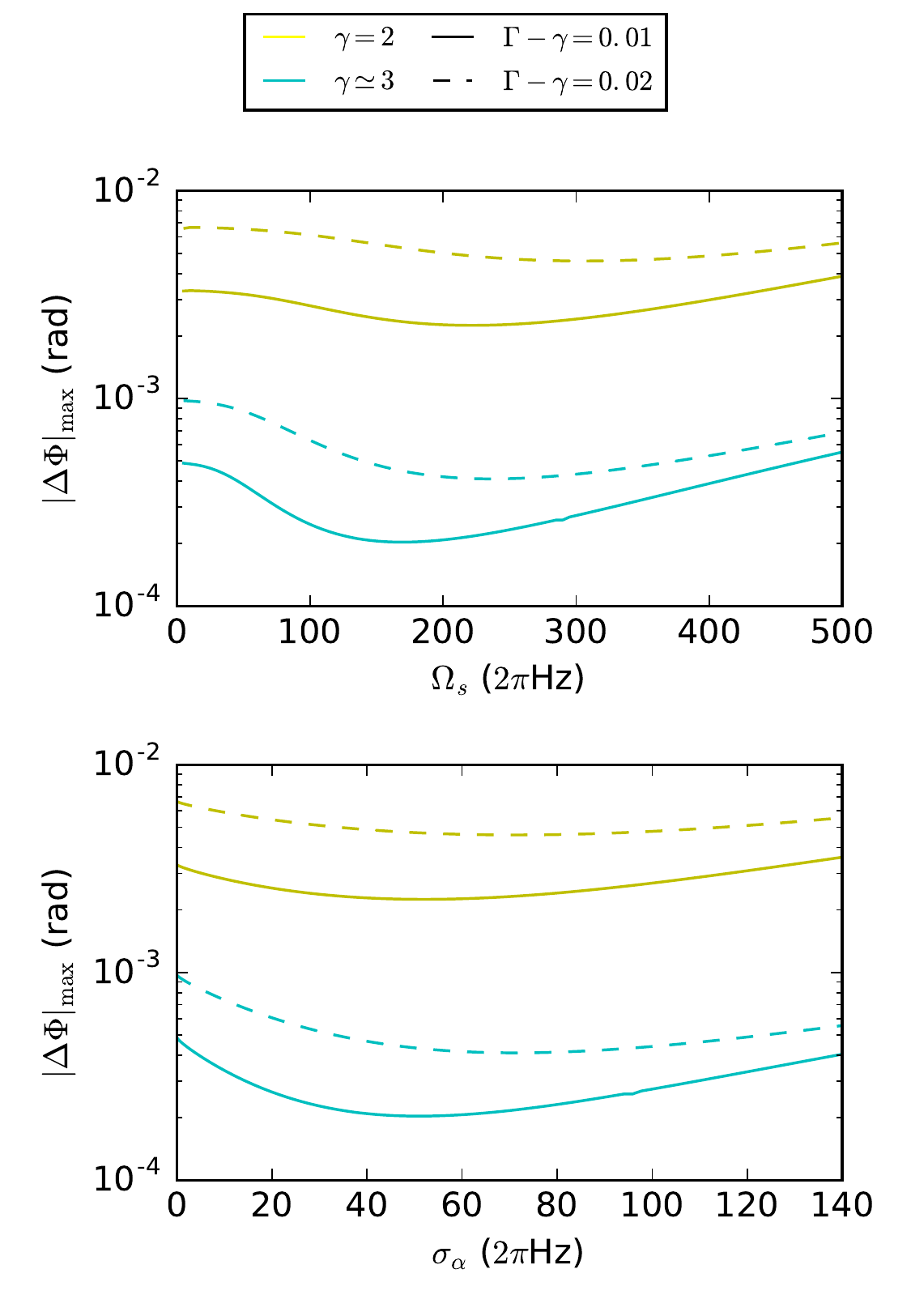}
\caption{Maximum GW phase shift $|\Delta\Phi|_{\rm max}$ for one of
  the $j=3,~m=1$ i-modes (the yellow curves in Fig.~\ref{m1_mix}) for
  different stellar models. The NS parameters are $M_{1.4}=R_{10}=1$.
    Upper panel: $|\Delta\Phi|_{\rm max}$ as a function of stellar
    spin $\Oms$. Lower panel: $|\Delta\Phi|_{\rm max}$ as a function
    of the inertial-frame frequency (which is equal to the GW
    frequency). Different color denotes different density models and
    the solid and dashed lines denote relatively low
    ($\Gamma-\gamma=0.01$) and high ($\Gamma-\gamma=0.01$)
    stratification, respectively.}
\label{special_i}
\end{figure}

\begin{figure}[!]
\centering
\includegraphics[width=.5\textwidth]{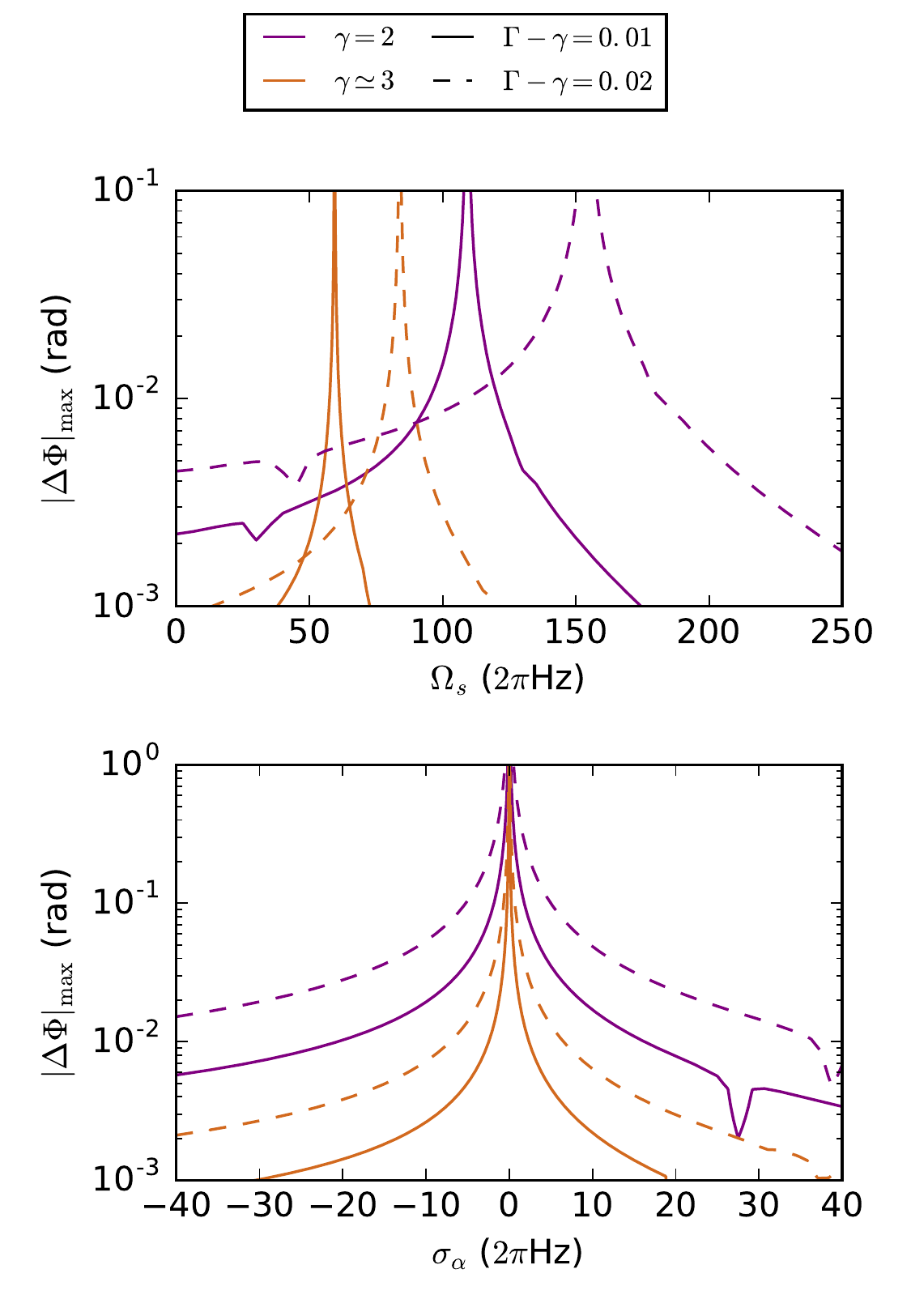}
\caption{Similar to Figure \ref{special_i}, but for the modified $m=2$ g-mode that 
has $\sigma\al$ zero crossing in frequency (purple curves in Fig.~\ref{m2_mix}). 
}
\label{special_g}
\end{figure}

\subsection{Pure Inertial Modes}

With finite NS rotation (but no stratification), we have
$\sigma_\alpha\propto \Omega_s$ and $Q_{\alpha,2,m}\propto \Omega_s^2$. 
Equation (\ref{eq:orbchange}) can be written as 
\begin{eqnarray}
&&\Delta\Phi=\mp 0.0027 \left({R_{10}^8\over M_{1.4}^6}\right){2\over q(1+q)}
\left({\epsilon_\alpha|\sigma_\alpha|\over\Omega_s^2}\right)^{\!-1}\nonumber\\
&&\qquad\quad \times \left({Q_{\alpha,2m}\over 0.02\hat\Omega_s^2}\right)^2
\left({f_s\over 500\,{\rm Hz}}\right)^2\left|{\cal D}^{(2)}_{m\pm 2}\right|^2,
\label{eq:imode-phase}\end{eqnarray}
where $f_s=\Omega_s/(2\pi)$ is the NS rotation frequency, 
the upper (lower) sign applies to modes with 
$\sigma_\alpha>0$ ($\sigma_\alpha<0$), which are excited by the $m'=2$
($m'=-2$) tidal potential. The relevant Wigner functions are
\begin{eqnarray}
&& |{\cal D}^{(2)}_{22}|=\cos^4{\Theta\over 2},\\
&& |{\cal D}^{(2)}_{2-2}|=\sin^4{\Theta\over 2},\\
&& |{\cal D}^{(2)}_{12}|=2\cos^3{\Theta\over 2} \sin {\Theta\over 2},\\
&& |{\cal D}^{(2)}_{1-2}|=2\sin^3{\Theta\over 2} \cos {\Theta\over 2},
\end{eqnarray}
where $\Theta$ is the spin-orbit misalignment angle.
From Table IV we see that $Q_{\alpha,2m}\lesssim 0.02{\hat\Omega_s}^2$, thus pure
i-modes give rise to a negligible $\Delta\Phi$ unless the star has $R_{10}^8/M_{1.4}^6\gg 1$.

\subsection{Mixed Modes}

In the presence of NS rotation and stratification, we write Eq.~(\ref{eq:orbchange}) 
(using Eqs.~\ref{eq:scaleomega} and \ref{eq:scaleQ}) in the form
\begin{eqnarray}
&&\Delta\Phi=\mp 0.060\left({R_{10}\over M_{1.4}}\right)^{\!5}{2\over q(1+q)}
\left({\Gamma-\gamma\over 0.01}\right)\nonumber\\
&&\qquad\quad\times \left(\!{{\bar f}_\alpha\over 100\,{\rm Hz}}\!\right)^{\!\!-2}
\left({{\bar Q}_{\alpha,2m}\over 10^{-3}}\right)^{\!2}
\left({2\pi f_\alpha\over\epsilon_\alpha}\right)\left|{\cal D}^{(2)}_{m\pm 2}\right|^2
\nonumber\\
&&\qquad = \mp 0.060\left({R_{10}^8\over M_{1.4}^6}\right){2\over q(1+q)}
\left({f_\alpha\over 100\,{\rm Hz}}\right)^2\nonumber\\
&&\qquad\quad\times \left(\!{{\bar f}_\alpha\over 100\,{\rm Hz}}\!\right)^{\!\!-4}
\left({{\bar Q}_{\alpha,2m}\over 10^{-3}}\right)^{\!2}
\left({2\pi f_\alpha\over\epsilon_\alpha}\right)\left|{\cal D}^{(2)}_{m\pm 2}\right|^2,
\label{eq:mixmode-phase}\end{eqnarray}
where $f_\alpha=|\sigma_\alpha|/(2\pi)$ is the absolute value of the mode frequency, 
and ${\bar f}_\alpha={|\bar\sigma}_\alpha|/(2\pi)$ and ${\bar Q}_{\alpha,2m}$
can be read off from Figs.~\ref{m1_mix}-\ref{m2_mix} for different NS models and modes.
In Eq.~(\ref{eq:mixmode-phase}), the upper (lower) sign applies to modes with 
$\sigma_\alpha>0$ ($\sigma_\alpha<0$), i.e., the prograde (retrograde) modes with respect to
the spin axis in the inertial frame, which are excited by the $m'=2$ ($m'=-2$) tidal potential.

Figure \ref{Phi_mixed} shows the GW phase shift at the optimal
inclination (i.e., for the value of $\Theta$ that maximizes $|{\cal D}^{(2)}_{m\pm 2}|$) 
for the modes depicted in Figs.~\ref{m1_mix}
and \ref{m2_mix}. Note that the phase shift of the $j=m=1$ r-mode (the
blue curves in Fig.~\ref{m1_mix}) is now shown here because its
frequency $\sigma_\alpha$ is essentially zero.  We see that for most
modes with canonical NS parameters ($M_{1.4}=R_{10}=1$), the GW phase
shift is much less than unity. In fact, with two exceptions (which we
will discuss in the next paragraphs), the GW phase shift of a mixed
mode is not significantly larger than that of the corresponding pure
g-mode when rotation is ignored, or the corresponding pure i-mode when
stratification is ignored.
This comes about because the tidal coupling $Q_{\alpha,2m}$ is usually not 
significantly enhanced by mode mixing. As a result, for these modes, knowing the pure g-mode and
pure i-mode results is enough to give a good estimate of the GW phase shift.

There are two cases where the combination of rotation and stratification significantly
enhances the GW phase shift, and an estimate of $\Delta\Phi$ based on pure i-modes
or pure g-modes would prove inadequate. The first case concerns the i-modes
that are significantly modified by stratification. An example of this
is the yellow mode shown in Fig.~\ref{m1_mix}. When $\Oms\to 0$, this mode
has $\sigma_\alpha\to 0$ and does not become any of the g-modes;
however, the mode frequency and tidal coupling are both significantly
affected by stratification (see Fig.~\ref{mixed_i}). As a result, we see
in Figure \ref{Phi_mixed} that $\Delta\Phi$ for this mode converges to a finite
value as $\Oms\to 0$. In particular, for a wide range of $\Omega_s$ (including 
$\Omega_s\to 0$), $\Delta\Phi$ of this mode 
is larger than all other modes shown in Fig.~\ref{Phi_mixed}. 
To study this mode in more details, we plot the
maximum GW phase shift as a function of $\Oms$ and $\sigma\al$ for
different stellar models in Figure \ref{special_i}. We see that 
$|\Delta\Phi|$ decreases as $\gamma$ increases; this agrees
with the fact that $Q_{\alpha,21}$ decreases when $\gamma$ increases
(see Table \ref{i_modes}). Also, $|\Delta\Phi|$ increases when $\Gamma-\gamma$
increases; we find that $|\Delta\Phi|\propto \Gamma-\gamma$ at $\Oms\to
0$, which is a scaling relation similar to pure g-modes. For larger
$\Oms$, stratification affects $\Delta\Phi$ less significantly. Moreover, this mode maintains a relatively large (compared to
other modes) and nearly constant (varying by $\lesssim 50\%$)
$\Delta\Phi$ as $\sigma\al$ varies from zero to a few hundred Hz,
while most other modes have zero $\Delta\Phi$ at $\sigma\al\to0$
(because for those modes small $\sigma\al$ requires either stratification or
rotation to be small).
Note that this mode has very small frequency ($\sigma\al\ll \Oms$) for
small $\Oms$; therefore it is necessary to consider whether rotational
distortion affects the above results. It turns out that rotational
distortion only modifies $\sigma\al$ and $\Delta\Phi$ slightly.

Another important case is when the inertial-frame frequency $\sigma\al$
becomes small due to the combined effects of stratification and rotation.
This occurs when a retrograde mode (with $\omega_\alpha<0$)
gets ``dragged'' by the NS rotation, so that $\sigma_\alpha$ increases and changes sign 
as $\Oms$ increases. As a result,
$\sigma\al$ becomes zero for a particular value of $\Oms$, which causes the GW
phase shift $\Delta\Phi$ to diverge (since $\Delta\Phi\propto
1/|\sigma\al|$).\footnote{It is in principle possible for $\Delta\Phi$
  to diverge due to $\epsilon\al$ crossing zero; but this never
  happens for the modes studied in this paper.} 
An example is the $m=2$ purple mode depicted in Fig.~\ref{m2_mix}, which has a negative $\sigma\al$
for $\Oms\to 0$ (``retrograde'') and positive $\sigma\al$ (``prograde'')
for large $\Oms$. We see from Fig.~\ref{Phi_mixed} that $|\Delta\Phi|_{\rm max}$ diverges at
$\Oms\simeq 110\cdot2\pi$Hz. 
Figure \ref{special_g} shows the maximum
GW phase shift due to this $m=2$ mixed mode as a function of $\Oms$ and $\sigma\al$ for different
NS models. We see that for larger $\gamma$, the zero crossing
($\sigma_\alpha=0$) happens at smaller $\Oms$ and the peak is narrower, while for larger
$\Gamma-\gamma$, the zero crossing happens at larger $\Oms$ and the
peak is wider. For $\gamma=2,\Gamma-\gamma=0.02$ (which gives the
largest $|\Delta\Phi|_{\rm max}$ among the four sets in Fig.~\ref{special_g}), 
we see that $|\Delta\Phi|_{\rm max}\gtrsim 0.01$~rad for a
$\sim 100\cdot 2\pi$Hz range in $\sigma\al$ (which corresponds to a $\sim 50$~Hz
range in the GW frequency). This relatively large width suggests that
there is an appreciable chance for the system to have a combination of
spin and stratification profile that gives a small enough $|\sigma\al|$
to cause a significant GW phase shift.

\subsection{Summary: GW Phase Shift for Different GW Frequencies}

In order to relate our results to potential future observations, we summarize
here the GW phase shifts we expect to see at different GW
frequencies. Note that the GW frequency ($f_{\rm GW}$) is equal to
the absolute value of the inertial-frame mode frequency (since we only
consider the tidal potential with $m'=\pm 2$), i.e. $f_{\rm GW}
=f_\alpha=|\sigma\al|/(2\pi)$. For convenience, we consider a NS with 
$M_{1.4}=R_{10}=1$ and an equal-mass companion, but results for other NS/binary
parameters can be obtained by appropriate scalings.

For $f_{\rm GW}\lesssim 20$~Hz, the most significant GW phase shift
comes from the tidal resonance of zero-crossing mixed modes (or modified
g-modes); see the purple-lined mode shown in Figs.~\ref{m2_mix} and \ref{Phi_mixed}) 
with small $\sigma\al$. These modes can have large GW phase shift (even
$\gtrsim 1$~rad), and the $|\Delta\Phi|$ is larger for smaller $\sigma\al$ 
(and thus smaller $f_{\rm GW}$).
Note that to have a small $\sigma_\alpha$ requires the NS to have a right
combination of stratification and rotation (see Fig.~\ref{special_g}).
For a given NS binary, the probability that the parameters of the system gives a
small $\sigma\al$ is relatively low. 

Another major source of GW phase shift at low frequency ($f_{\rm GW}\lesssim 100$~Hz) 
is the stratification-modified i-mode
(see the yellow-lined mode in Figs.~\ref{m1_mix} and \ref{Phi_mixed}; see also
Fig.~\ref{special_i}). For these modes the GW frequency is determined by the spin rate and
stratification together, but the phase shift is mainly determined by
the strength of stratification. As a result, there is little
correlation between phase shift and GW frequency. For typical
stratification ($\Gamma-\gamma\sim 0.01$), the phase shift is $\sim
10^{-3}$ to $10^{-2}$ rad.

For higher frequency ($f_{\rm GW}\gtrsim 100$~Hz), the most
significant GW phase shift can come from g-mode, i-mode or mixed
modes, depending on the parameters of the system. These modes share
the common features that the GW phase shift tends to increase as
$f_{\rm GW}$ increases, while $f_{\rm GW}$ tends to increase when
stratification and spin rate increase. NSs with large stratification
and/or spin rate will have modes at relatively high frequency (a few
hundred Hz) that give a relatively large ($\gtrsim 0.01$~rad) GW phase
shift. 

\section{Summary}

We have presented a comprehensive study on the resonant tidal
excitation of neutron star (NS) oscillation modes in coalescing
compact binaries. Such ``resonant tide'' may affect the gravitational
waveforms from the binary inspiral, and could potentially provide a
clean window for studying NSs using gravitational waves. Our study 
goes beyond previous works in that we treat the effects of NS rotation and
stratification exactly (using a newly developed spectral code for 
NS oscillations) -- this exact calculation reveals several features
of the resonant tide that are not present in previous works.
The main results of our paper are summarized as follows.
\begin{itemize}
\item Given the various uncertainties associated with the NS equation of state
and in preparation for future studies 
of NSs using GWs, we have adopted parameterized 
polytropic models that characterize the density and stratification profiles
of the NS. Such a parameterization provides a ``survey'' for various possible
NS models. Throughout the paper, we have presented scaling relations
for the NS oscillation mode frequency, tidal coupling coefficient and the GW phase shift 
associated with a resonance. Thus, while our numerical results and figures are often 
specific to a particular NS model, they can be rescaled when a different 
NS model is considered.

\item We have developed a new spectral code to calculate the oscillation modes
of rotating NSs, with an exact treatment of the Coriolis force. This exact
(non-perturbative) treatment allows us to obtain the mapping of various modes
as a function of the rotation rate (see Figs.~3-4).
Our spectral code can also include the effects of rotational distortion of the 
unperturbed star and the gravitational perturbation (i.e., without using the 
Cowling approximation). 
We find that although adopting the Cowling approximation barely affects
the mode frequency, it can cause appreciable overestimate
of the tidal coupling coefficient and the GW phase shift.
Overall, while these high-order effects (rotational distortion and gravitational perturbation)
do not qualitatively change the
tidal excitation property of most oscillation modes, there is one
important exception: We have shown that when all these effects are included,
the $j=m=1$ r-mode has essentially zero frequency, whereas approximate
calculations would give $\sigma_\alpha/\Omega_s \sim \hat\Omega_s^2$.
(see Section IV.B.4 for more details). This is important because it implies that 
the $j=m=1$ r-mode cannot be tidally excited during binary inspiral.

\item For pure g-modes (with negligible rotation), the mode frequency $f_\alpha$
(which is also the corresponding GW frequency at resonance)
depends on the stratification, and the GW phase shift $|\Delta\Phi|$ is always
$\lesssim 0.01$ for canonical NSs ($M=1.4M_\odot$, $R=10$~km) and equal-mass binaries
(Fig.~8). However, $\Delta\Phi$ increases as $R^8/M^6$
(Eq.~38) and can become significant for low-mass NSs with larger radii.
For pure inertial modes (with no stratification), the phase shift is typically smaller,
but it also increases with increasing $R^8/M^6$ (Eq.~39).

\item In the presence of both rotation and stratification, a NS has a
  spectrum of mixed (inertial-gravity) modes (Figs.~3-4), two of which
  may lead to appreciable GW phase shift at low frequency
  (Fig.~9). The first can be thought of as the rotation-modified $m=2$
  g-mode (the purple lines in Fig.~4 and Fig.~9; see also Fig.~11):
  this mode is retrograde in the rotating frame of the NS, but attains
  a small inertial-frame frequency $\sigma_\alpha$ because of
  rotation. A significant $\Delta\Phi$ can be produced when the NS has
  an appropriate rotation to give $f_\alpha=|\sigma_\alpha|/(2\pi)\lesssim 20$~Hz.
The second mixed mode of interest is the stratification-modified $m=1$ i-mode
(the yellow lines in Fig.~3 and Fig.~9; see also Fig.~10): This mode
has a frequency that approaches zero for $\Omega_s\rightarrow 0$, but is nevertheless
significantly affected by the stratification; as a result, $\Delta\Phi$ for this mode
is mainly determined by the stratification and does not depend sensitively on the GW
frequency (Fig.~10). The value of $\Delta\Phi$ for this mode
is still small ($\lesssim 10^{-2}$) for canonical NSs, but as for all the modes studied 
in this paper, $|\Delta\Phi|$ increases with increasing $R^8/M^6$
(Eq.~44).

\end{itemize}

\begin{acknowledgments} 
This work has been supported in part by NASA grant NNX14AP31G and a
Simons Fellowship in theoretical physics (DL).  WX acknowledges the
supports from the Hunter R. Rawlings III Cornell Presidential Research
Scholar Program and a special undergraduate research fellowship from
the Hopkins Foundation.
\end{acknowledgments}

\appendix
\newpage
\section{Numerical Methods}

Here we sketch our numerical methods used to solve for the oscillation
modes in a rotating star with a general density profile.  For
simplicity, we ignore the centrifugal distortion of the equilibrium
star in our discussion here, although this can be included using the
method discussed in Ref.~\cite{Reeseetal06}.

\subsection{Scaling of Equations}

When solving for the eigenmodes of the generalized eigenvalue problem 
\eqref{unscaled1} - \eqref{unscaled4}, it is convenient to use the normalization
\eq{
R  = 4\pi G = \rho_c = 1,
}
where $\rho_c$ is the central density of the star. We can define a scaled equilibrium density
\eq{
H(r) \equiv \left[\rho_0(r)\right]^{1/N},
}
where $N= 1/[\gamma(1)-1]$, 
and $\gamma(1)$ is the value of $\gamma$ evaluated at the stellar surface ($r=1$), with  
$\gamma=d\ln p_0/d\ln \rho_0$. 
We also scale the density and pressure perturbation $\delta\rho$ and $\delta p$ using
\eq{
b\equiv \delta\rho/H^{N-1},~~~ \Pi \equiv \delta p/H^N,
}
where we have assumed that $N\geq 1$. Such scaling guarantees the regularity of the 
solution at the stellar surface.

With the scaled variables, equations \eqref{unscaled1}-\eqref{unscaled4} can be written as
\ea{
&-i\omega b = - N\delta\bv\cdot\nabla H - H\nabla\cdot \delta\bv,\label{scaled1}\\
&-i\omega H\delta\bv  = -H(\nabla\Pi + \nabla\Phi)+\nabla H \left(-N\Pi + \frac{N\gamma}{N+1}\frac{b}{\Lambda}\right)\notag\\
&\phantom{-i\omega H\delta\bv  =}-2H\mathbf{\Oms}\times \delta\bv,\label{scaled2}\\
&-i\omega\left(\Pi - \frac{\Gamma}{(N+1)\Lambda}b\right) =\left(\frac\Gamma\gamma -1\right)\frac{N\gamma}{N+1}\frac{\delta\bv\cdot\nabla H}{\Lambda},\label{scaled3}\\
&0 = \Delta(\delta\Psi)-H^{N-1}b,\label{scaled4}
}
where
\eq{
\Lambda&\equiv \frac{\rho_0^{1+1/N}}{(N+1)p_0}.
}
Note that when the star is exactly polytropic, $\frac{N\gamma}{N+1}$
is always 1 and the above equations reduce to equations (22)-(24) in
\cite{Reeseetal06} (with $\lambda=-i\omega$).

Equations \eqref{scaled1}-\eqref{scaled4} are still in 2D; to further reduce the problem 
we remove the azimuthal dependence by expanding $\Pi,b$ and 
$\delta\Psi$ 
in terms of spherical harmonics:
\eq{
\Pi = \sum_{j=m}^{\infty} \Pi_{jm}Y_{jm},~~~ b = \sum_{j=m}^{\infty} b_{jm}Y_{jm},
~~~\delta\Psi = \sum_{j=m}^{\infty} \Psi_{jm}Y_{jm},
}
where $\Pi_{jm},~b_{jm}$ and $\Psi_{jm}$ are functions of $r$.
Similarly, the velocity perturbation $\delta\bv = -i\omega\bxi$ can be decomposed into 
spheroidal and toroidal components:
\eq{
\delta\bv = &\sum_{j=m}^{\infty}\Bigl[u_{jm}Y_{jm}\mathbf{e}_r + v_{jm}\left(\partial_\theta Y_{jm}\mathbf{e}_\theta + D_\phi Y_{jm}\mathbf{e}_\phi\right) \\
&+ w_{jm}\left(D_\phi Y_{jm}\mathbf{e}_\theta - \partial_\theta Y_{jm}\mathbf{e}_\phi\right)
\Bigr],
}
where $u_{jm},v_{jm},w_{jm}$ are functions of $r$,
$\mathbf{e}_r,\mathbf{e}_\theta,\mathbf{e}_\phi$ are unit vectors in
$r,\theta,\phi$ direction respectively, and $D_\phi\equiv
(\sin\theta)^{-1}\partial_\phi$. The terms with $u_{jm},v_{jm}$ give the spheroidal
component of the velocity perturbation and while the terms with $w_{jm}$ the toroidal
component. Thus, the mode is fully described by six sets of variables
($u_{jm},v_{jm},w_{jm},\Pi_{jm},\Psi_{jm},b_{jm}$) that only depend on
$r$.

\subsection{Solving the Equations}

To solve the eigenvalue problem, we transform \eqref{scaled1}-\eqref{scaled4} into 
equations of different $j$ components of the
variables $u,v,w,\Pi,\Psi,b$. We multiply Eqs.~\eqref{scaled1},
\eqref{scaled3} and \eqref{scaled4} by $\{Y_{jm}\}^*$ and integrate
over the $4\pi$ solid angle; this effectively projects the equations onto
different $Y_{jm}$. For \eqref{scaled2}, we project the equation onto
$(Y_{jm}\mathbf{e}_r)$, $(\partial_\theta
Y_{jm}\mathbf{e}_\theta+D_\phi Y_{jm}\mathbf{e}_\phi)$ and $(D_\phi
Y_{jm}\mathbf{e}_\theta-\partial_\theta Y_{jm}\mathbf{e}_\phi)$
respectively (by multiplying the equation by the conjugate of the projection vector
and integrate over $4\pi$ radians). For details of this projection and the
expression of the projected equations \footnote{The projected equations
  in Ref.~\cite{Reeseetal06} are somewhat different from ours since
  they assume a strictly polytropic star, but the method of obtaining the
  equations is similar.}, see
Ref.~\cite{Reeseetal06}. Then we choose a cutoff $j$ (call it $\jm$) and
keep only components with $j\leq\jm.$ For the $r$ direction, we choose
a Chebychev grid consisting of $N_r$ points from 0 to 1, which allows
us to use spectral method to solve for the modes. The problem is now
reduced to a generalized eigenvalue problem of matrices. Note that the
use of spectral method requires that the stellar profile (density,
stratification, etc.) to be relatively smooth.

For implementation of the boundary conditions, see \cite{Reeseetal06}.

\subsection{Limitations of the Method}

A major limitation of the method is that we must assume $N\geq 1$ when
setting the scaled variables. As a result, this method does not work
for stellar models with $N<1$. Moreover, in practice when $N$ is not
an integer the accuracy of the algorithm drops significantly, possibly
due to the fact that we are using a spectral method which
effectively expands the mode into basis polynomials.

Fortunately, there is an easy way to fix this problem. Since we do not
require a constant $\gamma$ throughout the star (unlike
\cite{Reeseetal06}), we can always modify $\gamma$ near the outer
boundary so that most of the star has the density profile we desire,
and towards the boundary $\gamma$ goes to a value which gives a
positive integer $N$ and thereby ensuring good performance of our
algorithm. In this way, we can obtain a reasonably good approximation
for any density profile we need. One example is the $\gamma\simeq 3$
model used in the main text (Section IV).

\bibliography{BIB.bib}

\end{document}